\PassOptionsToPackage{dvipsnames}{xcolor}
\documentclass[aps,prb,reprint,superscriptaddress,showpacs,notitlepage,floatfix]{revtex4-2}
\usepackage[T1]{fontenc}
\usepackage[utf8]{inputenc}
\usepackage[american]{babel}

\usepackage{amsfonts}
\usepackage{amsmath}
\usepackage{amssymb}
\usepackage{bm}
\usepackage{booktabs}
\usepackage{graphicx}
\usepackage{hyperref}
\usepackage{latexsym}
\usepackage{mathtools}
\usepackage{soul}
\usepackage{subfig}
\usepackage{xcolor}
\usepackage{xfrac}
\usepackage[normalem]{ulem}

\usepackage{gnuplot-lua-tikz}
\usepackage{tikz}
\usetikzlibrary
  {arrows,calc,through,backgrounds,matrix,positioning,decorations.pathmorphing}

\renewcommand{\vec}[1]{\boldsymbol{#1}}

\newcommand{\bB}{\boldsymbol B}

\newcommand{\im}{\mathrm{i}}

\newcommand{\rep}{\mathrm{Re}}
\newcommand{\imp}{\mathrm{Im}}

\begin{document}

\title{Dyakonov-Shur instability of electronic fluid: Spectral effect of weak magnetic field}

\date{Draft of \today}

\author{Matthias Maier}
\thanks{\url{maier@tamu.edu}; \url{https://people.tamu.edu/~maier}}
\affiliation{Department of Mathematics, %
  Texas A\&M University, College Station, Texas 77843, USA}

\author{Dennis Corraliza-Rodriguez}
\thanks{\url{dc858@cornell.edu}; %
  \url{https://www.cam.cornell.edu/research/grad-students/dennis-corraliza}}
\affiliation{Center for Applied Mathematics, %
  Cornell University, Ithaca, New York 14853, USA}

\author{Dionisios Margetis}
\thanks{\url{diom@umd.edu}; \url{https://www.math.umd.edu/~diom}}
\affiliation{Institute for Physical Science and Technology, %
  and Department of Mathematics, University of Maryland, College %
  Park, Maryland 20742, USA.}

\begin{abstract}
  We study numerically and analytically how the Dyakonov-Shur instability
  for a two-dimensional (2D) inviscid electronic fluid in a long channel
  can be affected by an external, out-of-plane static magnetic field. By
  linear stability analysis for a model based on the shallow-water
  equations, we describe the discrete spectrum of frequencies. When the fluid system is near the subsonic-to-supersonic
  transition point, a magnetically controlled gap between the stability and
  instability spectra of the complex eigen-frequencies is evident by our
  computations. This suggests that,  within this model,   the passage from stability to
  instability (and vice versa) is no longer continuous in the effective
  Mach parameter of the boundary conditions.  We also demonstrate that the growth exponents are enhanced by the magnetic field.   In a regime of weak magnetic fields, we derive a scaling law for
  the eigen-frequencies by perturbation theory. We discuss  theoretical   implications of
  our results in efforts to generate terahertz electromagnetic radiation by
  2D electronic transport.
\end{abstract}

\maketitle


\section{Introduction}

For the last two decades, the problem of efficiently generating
electromagnetic radiation at terahertz (THz) frequencies has received
considerable attention~\cite{Tonouchi2007, Kitaeva2008, Rana2019,
Dhillon2017, Sengupta2018, Fullop2020}. Research in this direction is
motivated by a plethora of potential applications, which include the
identification of distinct chemical processes, diagnostics for industrial
and environmental purposes, biomedical sensing and imaging, and the
monitoring of microscopic dynamics in semiconductors and
nanomaterials~\cite{Kitaeva2008}. THz waves are technologically appealing
since they may attenuate slowly with wavelengths suitable for nanoscale
sensing~\cite{Chan2007,Woolard2005,Hattori2007}. Experimental designs for
the generation of THz waves aim to capture motions of microscale dynamics,
e.g., vibrations of large molecules and electronic
transitions~\cite{Knoesel2001,Rana2007,Hauri2011}.

At the same time, there are intensive efforts to harness properties of
two-dimensional (2D) materials such as graphene and van der Waals
heterostructures~\cite{Torres2014, Geimetal2013, CastroNetoetal2009,
Novoselovetal2012}. When the size of the sample is large compared to the
mean free path for momentum-conserving electron-electron scattering but
small compared to the mean free path of momentum-relaxing collisions, the
electron system may behave as a fluid~\cite{Gurzhi1968, LucasFong2018}.
This behavior has been observed experimentally; see
e.g.~\cite{Bandurin2016negative, Crossno2016observation,
Kumar2017superballistic, Bandurin2018}. Accordingly, the electronic density
and velocity fields can be described by hydrodynamic
approaches~\cite{LucasFong2018}.

Three decades ago, in a seminal paper Dyakonov and Shur (DS) formulated a
one-dimensional (1D) model based on the shallow-water equations in their
proposal for plasma wave generation in a field effect
transistor~\cite{DyakonovShur1993}. A key ingredient of their model is its
boundary conditions with fixed electronic surface density and normal flux
at the edges of a long channel. By linear stability analysis, this model
leads to the prediction of an instability in the subsonic
regime~\cite{DyakonovShur1993}; and has inspired the attractive idea of
generating THz waves via unstable 2D electronic
transport~\cite{LucasFong2018}. In the DS theory, the passage from
 fluid stability  to instability can occur   by a
continuously varying Mach number formed by the intrinsic sound speed and a
parameter from the boundary conditions.

To our knowledge, the realization of the DS instability in the laboratory
setting remains an alluring yet challenging and elusive goal to
date~\cite{Mendl2021, Crabb2022}. Research efforts to achieve this goal
have improved our understanding of how THz charge density waves can be
excited and sustained in 2D channels~\cite{Bhardwaj2016, Nafari2018,
Ryzhii2020, Shur2021}. Furthermore, various extensions of the DS model have
incorporated more realistic effects such as
nonlocality~\cite{DyakonovShur2005}, imperfection of the boundaries at the
ballistic-to-hydrodynamic crossover~\cite{MendlLucas2018}, and nonlinear
effects~\cite{Svintsov2013, Mendl2021}. In principle, the DS instability
can be significantly affected by details of the boundary conditions at the
crossover regime~\cite{MendlLucas2018}. In the hydrodynamic limit, the
character of this instability is undistorted; the related growth exponents,
however, are expected to be reduced by viscous
dissipation~\cite{LucasFong2018}.

In this paper, we study an extension of the DS model by including an
external, out-of-plane static magnetic field, $\bB$, in the Lorentz force.
The DS boundary conditions  for the density and normal flux are left
unchanged.   We also investigate two distinct choices of the boundary
condition for the tangential flux, for comparison purposes.  
The magnetic field breaks the time reversal symmetry of the shallow-water
equations formulated by DS~\cite{DyakonovShur1993}. By linear stability
analysis, we demonstrate how the resulting spectrum differs from the one
predicted by DS. Notably, the discrete eigen-frequencies of the requisite
non-Hermitian problem no longer lie along a straight line in the complex
plane. Near the subsonic-to-supersonic transition point, the complex point
spectrum exhibits a \emph{gap} between the imaginary parts of the
eigen-frequencies   in the stability and instability regimes.
 Such a gap   exists for each of our choices of the
boundary condition involving the tangential flux.   This
finding suggests that it would be  theoretically  
impossible to continuously connect   these two regimes   by
varying the Mach number of the DS model,   if the electronic
system could be driven in these regimes. A major underlying assumption is
the validity of the shallow-water equations of our model. For a
  weak magnetic field, we describe how  the
predicted   gap scales with $|\bB|$. Our analysis yields the
exact 2D electronic steady state,  which in principle depends on the
prescribed boundary conditions,   via the solution of a
nonlinear equation. We discuss implications of our results in efforts to
produce THz radiation.

Our work focuses on  exposing generic features of a particular
model, based on the Euler equations, for the hydrodynamic regime of 2D
electronic systems. We show   how a small perturbation of the
DS model~\cite{DyakonovShur1993}, by use of an external static magnetic
field, $\bB$, can cause qualitatively significant changes to
 related predictions.   For fixed nonzero $\bB$, the
electronic steady state is characterized by spatially varying density and
flux profiles, in contrast to the constant density and flux
in~\cite{DyakonovShur1993}. This steady state is locked into one of two
different branches   depending on the kinetic regime, or Mach
number, of the unperturbed DS model. Fluctuations of the fluid variables
around the steady state are bound to a particular branch; and exhibit a
stability or instability spectrum. By our numerics, we identify a
magnetically controlled, complex-valued spectral gap, in the sense defined
below (Sec.~\ref{sec:numerical_results}), near the transition point. This
gap is related to the multivalued function of the steady-state solution.

It is worthwhile to pose the following question: Can our analysis be useful
for future efforts to realize instabilities of 2D electronic fluids, or is
it merely an academic exercise? We tend to adopt the former point of view
here. We believe that our results offer a cautionary paradigm
 for modeling electron hydrodynamics,   since they
suggest how a magnetic perturbation of the DS model~\cite{DyakonovShur1993}
can lead to  modified predictions such as enhanced growth
exponents and a gapped complex-valued spectrum,   which might
challenge physical interpretation. We attribute this difficulty to the
nature of the underlying   hydrodynamic model and its boundary
conditions.    In particular, we discuss how the boundary
condition for the tangential flux affects the steady state and the fluid
spectrum.   The issue of whether the DS instability is
experimentally feasible or not is left unresolved.  We indicate
ranges of values of physical parameters that a 2D system may have to be a
plausible testbed for our predictions. However, we are unable to propose a
specific material for this purpose. 

From a purely theoretical perspective, we investigate a mechanism of time
reversal symmetry breaking of the unperturbed homogeneous system of
equations used by DS~\cite{DyakonovShur1993}. To our knowledge, this
extension of the DS model has not been previously explored, and paves the
way to the study of a broad family of effective hydrodynamic theories
lacking basic symmetries. From an applied perspective, we show that the
nonzero magnetic field, $\bB$, can plausibly enhance the growth exponents.

Our linear stability analysis largely relies on numerical computations.
Technically speaking, in our approach we discretize the differential
operator of the requisite non-Hermitian eigenvalue problem for the
frequencies by a finite difference approximation scheme. This method
enables us to numerically solve the differential equations for the
electronic surface density and flux. We compute a large number of the
(discrete) complex eigen-frequencies in the  fluid stability and
instability   regimes, and observe their behavior as a function
of the $\bB$ field and the Mach number of the DS model. The $\bB$ field
induces curvature to the continuous curves that interpolate the computed
eigen-frequencies. This situation should be contrasted to the case with
$\bB=0$ of the DS model, when the respective
eigen-frequencies lie in straight lines. By our numerics, we notice that as
$|\bB|$ increases, the instability spectrum is shifted towards higher
growth for fixed Mach number.

The remainder of this paper is summarized as follows.
Section~\ref{sec:background} provides a review of the DS model. We describe
the geometry, equations of motion and boundary conditions
(Sec.~\ref{subsec:DS-eqs}); outline the result of the linear stability
analysis (Sec.~\ref{subsec:DS-stabil}); discuss the sense in which the
  fluid stability and instability   spectra are
connected (Sec.~\ref{subsec:dist-spectra}); and point out a scaling limit
(Sec.~\ref{subsec:DS-scaling}). In Sec.~\ref{sec:model}, we describe the
hydrodynamic model with an out-of-plane static magnetic field, $\bB$: We
rescale and nondimensionalize the governing equations
(Sec.~\ref{subsec:rescaling}); and derive the electronic steady state via
solving an algebraic equation   for two distinct choices of the
boundary condition for the tangential flux  
(Sec.~\ref{subsec:steady_state}). Section~\ref{sec:linear_stability}
focuses on: the corresponding motion laws for a linear stability analysis
with a nonzero $\bB$ field  (Sec.~\ref{subsec:stabil-magn-govern-laws});
and a scaling prediction for the perturbed eigen-frequencies when the
$\bB$ field is weak (Sec.~\ref{subsec:stabil-magn-weakB}). In
Sec.~\ref{sec:numerical_results}, we carry out numerical simulations
(Secs.~\ref{subsec:spectrum} and~\ref{subsec:gap}), and discuss their
implications (Sec.~\ref{subsec:discussion}). Section~\ref{sec:conclusion}
provides an overlook and outline of open problems.

\noindent\emph{Notation and terminology}.
Throughout the paper, we write $f=\mathcal O(g)$ for scalars $f$ and $g$ to
mean that $|f/g|$ is bounded by a strictly positive constant in a
prescribed limit.  We use the term fluid
instability spectrum{--and fluid stability spectrum--}to
mean the discrete set of frequencies coming from the linear stability
analysis of the governing equations.


\section{Review of DS model}
\label{sec:background}

In this section, we review the main ingredients of the DS
model~\cite{DyakonovShur1993}. In particular, we describe the problem
geometry, equations of motion, and the associated boundary conditions; see
Sec.~\ref{subsec:DS-eqs}. In addition, we review the main result of the
linear stability analysis (Sec.~\ref{subsec:DS-stabil}); and point out a
scaling limit (Sec.~\ref{subsec:DS-scaling}).

\subsection{Geometry and governing equations}
\label{subsec:DS-eqs}

\begin{figure}
  \includegraphics*[scale=0.31, trim=0.9in 0.15in 0in 0.8in]{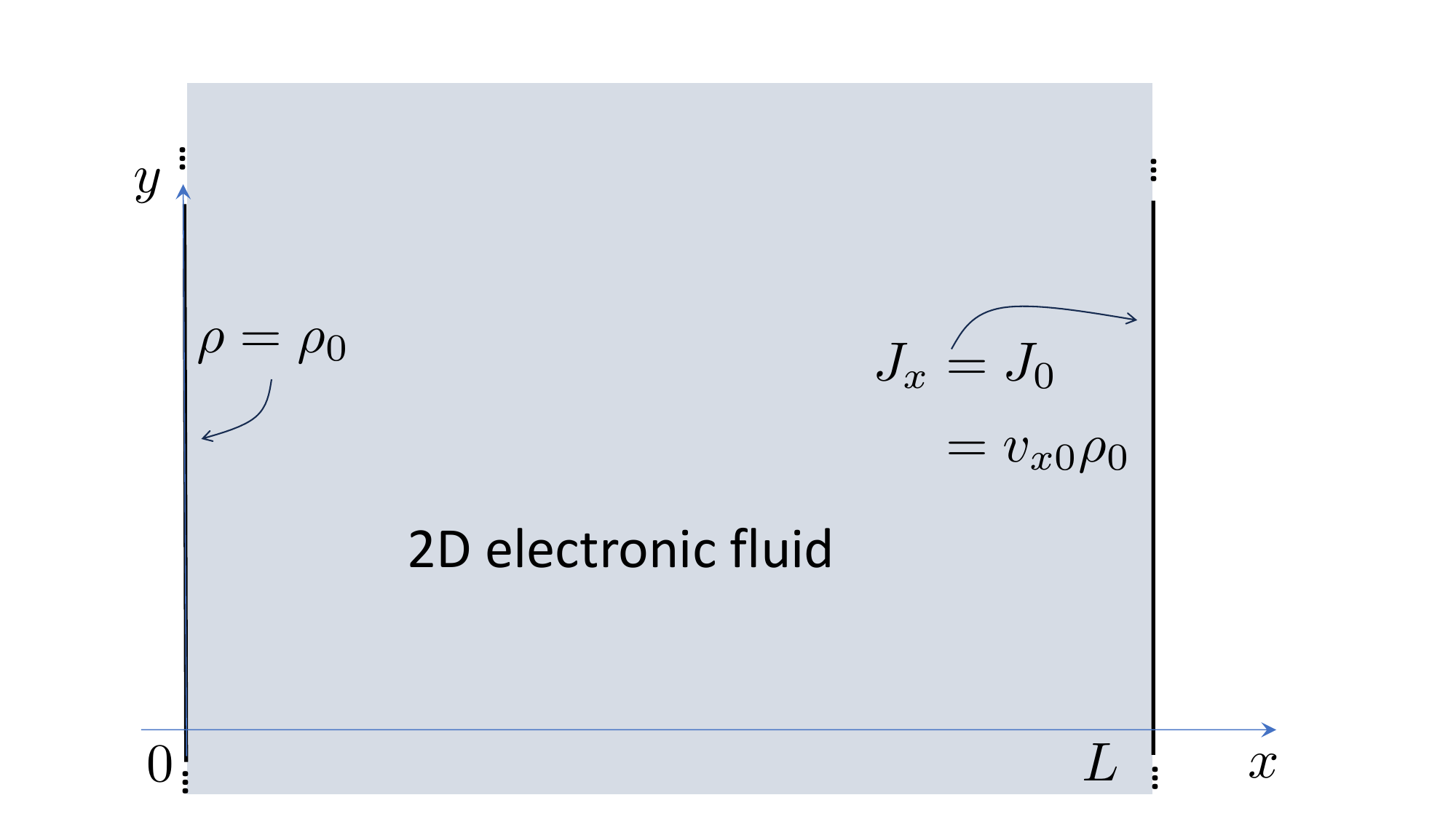}
  \caption{Geometry of the problem. The electronic fluid lies in an
    infinite-in-$y$ channel of width $L$, with edges at $x=0$ and $x=L$.
    The electronic density $\rho$ is fixed at the constant value $\rho_0$
    at the left boundary, $x=0$; and the normal flux $J_x$ has the constant
    value $J_0=v_{x0} \rho_0$ at the right boundary,
    $x=L$.}
  \label{fig:Geom}
\end{figure}

The problem geometry is shown in Fig.~\ref{fig:Geom}. The electronic fluid
occupies an infinitely long channel. The spatial coordinates are $(x, y)$
with  $0<x<L$ and $-\infty< y< +\infty$. The fluid flows in the
$x$-direction. The electronic number density per unit area is $\rho$, and
the vector-valued fluid velocity is $(v_x, v_y)$ with $v_y=0$. The
variables $\rho$ and $v_x$ are $y$-independent, and satisfy the system of
equations~\cite{DyakonovShur1993}
\begin{subequations}
\label{eqs:shallow_water_model_wo_B}
\begin{align}
  \left(\frac{\partial }{\partial t} +
  v_x \frac{\partial }{\partial x}\right)v_x
  \;&=\; - a \frac{\partial \rho}{\partial x}~, \label{eq:DS-velocity}
  \\
  \frac{\partial \rho}{\partial t} + \frac{\partial}{\partial x}(\rho\,v_x)
  \;&=\; 0,\quad 0<x<L, \label{eq:DS-mass-conserv}
\end{align}
where $a=e^2/(Cm_e)$ is a positive constant ($a>0$); $e$ is the electron
charge ($e>0$) and $C$ is a quantity expressing the effective capacitance
per unit area;  $m_e=\bar m_e m_{e0}$ is the effective
electron mass, and $m_{e0}$ is the free electron mass ($\bar
m_e=m_e/m_{e0}$).   This model comes from the Euler momentum
equation and the mass conservation law, where the Lorentz force is $\vec
F=(F_x, F_y)$ with $F_y=0$ and
\begin{equation}
  F_x=- e\, \frac{\partial U}{\partial x},\quad U\simeq \frac{e \rho}{C};
\end{equation}
\end{subequations}
and $U(t,x)$ is the electrostatic potential generated by the electronic
charge in the channel. The ``\emph{gradual channel approximation}'' is
applied here, by which the potential $U(t,x)$ is locally proportional to
the density $\rho(t,x)$~\cite{Shockley1952}.   We stress that, in
this 1D model, $v_y=0$ everywhere.  

In addition, the density $\rho$ and $x$-directed flux $J_x=\rho v_x$ obey
the boundary conditions (see Fig.~\ref{fig:Geom})
\begin{align}\label{eq:DS-BCs}
  \begin{rcases*}
    \begin{aligned}
      \rho&=\rho_0=\text{const}\quad \mbox{at}\ x=0,\\
      J_x&=v_{x0}\rho_0=\text{const}\quad \mbox{at}\ x=L,
     \end{aligned}
  \end{rcases*}
\end{align}
where $\rho_0$ and $v_{x0}$ are given positive constants ($\rho_0,
v_{x0}>0$). Thus, the fluid density is kept fixed to a constant at the left
boundary ($x=0$), while the normal flux is fixed to a constant at the right
boundary ($x=L$). DS  showed that   in this setting the sound waves either
decay or grow with time as they undergo multiple reflections at the two
boundaries~\cite{DyakonovShur1993,LucasFong2018}.

\subsection{Linear stability analysis}
\label{subsec:DS-stabil}

Let us discuss the particulars of the DS instability.
Equations~\eqref{eqs:shallow_water_model_wo_B} and~\eqref{eq:DS-BCs}
trivially admit the \emph{uniform} steady-state solution $(v_x,
\rho)=(v_{x0}, \rho_0)$. We perturb the fluid system around this solution
by setting~\cite{DyakonovShur1993}
\begin{equation*}
  \rho(t,x)=\rho_0+\rho_1(x) e^{-\Lambda t},\quad J_x(t,x)=\rho_0 v_{x0}+
  \jmath_{1}(x) e^{-\Lambda t},
\end{equation*}
where the functions $\rho_1$ and $\jmath_{1}$ as well as the parameter
$\Lambda$ should be determined consistently with the governing equations.
In the notation employed by DS~\cite{DyakonovShur1993}, $\Lambda=\im\omega$
where $\omega$ is the angular frequency (and $\im^2=-1$). By linearizing
the ensuing equations of motion for $(\jmath_{1}, \rho_1)$ we find an
eigenvalue problem for $-\im\Lambda=\omega$, which is expressed in a matrix
form as
\begin{subequations}\label{eqs:DS-sys}
\begin{equation}
  -\im \mathbb{A}_0\,\frac{\partial}{\partial x}
  \begin{pmatrix}
    \jmath_1 \\ \rho_1
  \end{pmatrix}
  =\omega
  \begin{pmatrix}
    \jmath_1\\ \rho_1
  \end{pmatrix},\quad 0<x<L,
\end{equation}
under the homogeneous boundary conditions $\rho_1(0)=0$ and
$\jmath_{1}(L)=0$. In the above, $\mathbb{A}_0$ is the $2\times 2$ matrix
\begin{equation}
  \mathbb{A}_0=
  \begin{pmatrix}
    2 v_{x0} & s_0^2-v_{x0}^2 \\
    1 & 0  \\
  \end{pmatrix},
\end{equation}
\end{subequations}
where $s_0=\sqrt{a \rho_0}$ is the sound speed ($s_0>0$). We refer to the
ratio $M=v_{x0}/s_0$ as the Mach number (or Mach parameter) of the DS
model.

The differential operator for the above eigenvalue problem for the
frequency $\omega$ is non-Hermitian. Thus, the eigen-frequencies
$\omega=-\im \Lambda$ can be complex. Indeed, a direct calculation
furnishes~\cite{DyakonovShur1993}
\begin{subequations}
\begin{equation}
  \label{eq:DS-eigenv-instab}
  \omega=\omega_n=-\im\frac{v_{x0}^2-s_0^2}{2Ls_0}\left[\ln\biggl(
    \frac{v_{x0}+s_0}{v_{x0}-s_0}\biggr)+\im 2n\pi\right].
\end{equation}
Here, $n$ is any integer ($n\in \mathbb{Z}$) and the logarithmic function,
$\ln(\cdot)$, is allowed to take complex values.

By Eq.~\eqref{eq:DS-eigenv-instab} we see that in the supersonic regime,
i.e., for $v_{x0}> s_0$, we have $\rep\,\Lambda>0$ which implies a stable
system~\cite{DyakonovShur1993}. In contrast, the fluid is unstable in the
subsonic regime ($v_{x0}<s_0$). In particular, by analytic continuation of
the $\ln(\cdot)$ function in Eq.~\eqref{eq:DS-eigenv-instab} we obtain
\begin{equation}
  \label{eq:DS-eigenv-stab}
  \omega_n=\im \frac{s_0^2-v_{x0}^2}{2Ls_0}\left[\ln\biggl(
  \frac{s_0+v_{x0}}{s_0-v_{x0}}\biggr)+\im (2n+1)\pi\right];
\end{equation}
\end{subequations}
thus, $\rep\,\Lambda<0$ if $v_{x0}<s_0$. In each regime the
eigen-frequencies $\omega_n$, viewed as points in the complex
$\omega$-plane,  lie in a straight line parallel to the real axis. In the
subsonic case, this line is in the upper half $\omega$-plane; whereas in
the supersonic case the line is in the lower half $\omega$-plane. We refer
to the point set $\{\omega_n\}_{n\in\mathbb{Z}}$ as the DS spectrum.

\subsection{Distance between DS spectra}
\label{subsec:dist-spectra}

It is worthwhile to define a measure of how far or close the  
fluid instability and stability   spectra are within the DS
model. This notion of a ``spectral distance'' can be used to estimate the
effect of kinetics, and can be extended to   the case with a
nonzero magnetic field.  

In the supersonic regime, $\omega_0$ is the eigen-frequency that is closest
to the origin of the $\omega$-plane. Let us introduce the convenient
dimensionless parameter
\begin{equation*}
  \delta=\frac{1}{2}\left(\frac{v_{x0}^2}{s_0^2}-1\right)=\tfrac{1}{2}(M^2-1);
\end{equation*}
thus, $\delta>0$ for stability. Accordingly, we can write the lowest
eigenvalue of  the stability spectrum, in the supersonic regime, as
\begin{subequations}
\begin{equation}
  \label{eq:lowest-eigenv-DS}
  \omega_0^{\text{DS}}(\delta)=-\im \frac{s_0}{L}\delta
  \ln\biggl[\frac{(1+\sqrt{1+2\delta})^2}{2\delta} \biggr],\ \delta>0.
\end{equation}
In this context, the switch from the supersonic to the subsonic regime can
be presented via the replacement of $\delta$  by $-|\delta|$. By analytic
continuation of the logarithm in the above expression for
$\omega_0^{\text{DS}}(\delta)$, we obtain the lowest eigen-frequencies of
the   fluid instability spectrum,    viz.,
\begin{equation}
  \label{eq:lowest-eigenv-DS}
  \omega_{0}^{\text{DS}}(-|\delta|)=\im \frac{s_0}{L}|\delta|
  \left\{\ln\biggl[\frac{(1+\sqrt{1-2|\delta|})^2}{2|\delta|} \biggr]\pm
  \im\pi\right\},
\end{equation}
where the double-valuedness is due to the unspecified branch for the
logarithm. When $\delta$ becomes sufficiently  small in this expression,
the imaginary part dominates and the ambiguity tends to disappear. We can
now define the distance, $D_\omega^{\text{DS}}$, between the  
fluid stability and instability spectra   as a function of
the parameter $\delta$ by
\begin{align}\label{eq:spectral-dist-def-DS}
  D_\omega^{\text{DS}}(\delta)&=|\omega_0^{\text{DS}}(|\delta|) -
  \omega_{0}^{\text{DS}}(-|\delta|)|\notag
  \\
  &\simeq 2\frac{s_0}{L}|\delta|\,
  \ln\biggl[\frac{(1+\sqrt{1+2|\delta|})(1+\sqrt{1-2|\delta|})}{2|\delta|}
  \biggr].
\end{align}
\end{subequations}
The first line is unambiguous since the result for
$D_\omega^{\text{DS}}(\delta)$ is the same regardless of the sign choice in
Eq.~\eqref{eq:lowest-eigenv-DS}. In the approximation of the second line of
Eq.~\eqref{eq:spectral-dist-def-DS}, we neglect the $\pm \im\pi$
contribution from Eq.~\eqref{eq:lowest-eigenv-DS}. Note that it suffices to
consider $\delta>0$ for the domain of $D_\omega^{\text{DS}}(\delta)$.

It is tempting to interpret formula~\eqref{eq:spectral-dist-def-DS}. In particular, we have $D_\omega^{\text{DS}}(\delta)\to 0$ as
$\delta\to 0$. We conclude that the passage from stability to instability
in the DS model is continuous in the Mach number
$M=v_{x0}/s_0$,   for any system that can exist in both regimes within this hydrodynamic model.   In other words, the transition from the lowest eigenvalue
of the   fluid stability spectrum   to its
counterpart of the   instability spectrum   happens
continuously with $M$.

\subsection{Scaling limit}
\label{subsec:DS-scaling}

Next, we outline in passing a scaling limit that converts the discrete DS
spectrum to a continuous one near the transition point, when
$|s_0-v_{x0}|/s_0$ approaches zero. By inspection of
Eq.~\eqref{eq:DS-eigenv-instab}, we define the wavenumber variable
\begin{equation*}
  k_n=\frac{(\rep\,\omega_n)}{v_{x0}+s_0}=\frac{v_{x0}-s_0}{s}\,\frac{n\pi}{L},
\end{equation*}
which has the dimension of inverse length. Note that the plane wave $e^{\im
k_n x}$ describes the spatially slow variation of the fluid density and
flux~\cite{DyakonovShur1993}. Evidently, $|k_{n+1}-k_n|L$ is small if
$|v_{x0}-s_0|/s_0 \ll 1$. Thus, $k_n=k$ tends to form a continuum of real
values in the fluid channel ($0<x<L$).

Consider the dimensionless complex frequencies
\begin{align}\label{eq:scaled-omega-DS}
  \tilde \omega_n = \frac{L \omega_n}{v_{x0}+s_0}= k_nL\biggl\{
  1-\frac{\im}{2n\pi}  \ln\biggl(\frac{2n\pi +k_nL}{k_nL}\biggr)\biggr\}.
\end{align}
In the asymptotic limit with $n\gg 1$ and $k_n L=\mathcal O(1)$, we can use
the approximation $\tilde\omega_n\simeq k_n L$. This limit suggests the
continuous real valued frequency spectrum
\begin{equation}\label{eq:DS-omega-k}
  \omega(k)=(v_{x0}+s_0)k=s_0(M+1) k,
\end{equation}
which is a linear dispersion relation for the sound waves in this system.


\section{Fluid model with magnetic field}
\label{sec:model}

In this section, we describe the governing hydrodynamic equations with a
magnetic field, and determine the corresponding steady state analytically.
  We investigate two distinct cases of the boundary condition for
the tangential flux.    We should emphasize that a major
difference of this setting with the  DS model~\cite{DyakonovShur1993} is
that the resulting steady state now becomes spatially dependent.

Consider the geometry of Fig.~\ref{fig:Geom}. The electronic fluid is
subjected to the out-of-plane, static magnetic field $\vec B = B \vec e_z$
where $B$ is a constant and $\vec e_z$ is the $z$-directed Cartesian unit
vector.   The equations of motions form an extension of
Eqs.~\eqref{eq:DS-velocity} and~\eqref{eq:DS-mass-conserv} to the setting
with the full 2D velocity vector field $\vec v=(v_x, v_y)$ and the Lorentz
force $\vec F=e(-\nabla U+\vec v\times \vec B)$ under the gradual channel
approximation~\cite{Shockley1952}. The shallow-water equations become
\begin{align}
  \label{eq:shallow_water_model_with_B}
  \begin{rcases*}
    \begin{aligned}
      \left(\frac{\partial}{\partial t} +
      v_x\frac{\partial}{\partial x} + v_y\frac{\partial}{\partial y} \right)v_x
      \;&=\; - a \frac{\partial\rho}{\partial x} + \frac{e}{m_e}\,v_y\, B,
      \\
      \left(\frac{\partial}{\partial t} +
      v_x\frac{\partial}{\partial x} + v_y\frac{\partial}{\partial y}\right)v_y
      \;&=\; - a \frac{\partial \rho}{\partial y} - \frac{e}{m_e}\,v_x\, B,
      \\
      \frac{\partial \rho}{\partial t}+ \frac{\partial}{\partial x}(\rho\,v_x)
      + \frac{\partial}{\partial y}(\rho\,v_y) \;&=\; 0.
    \end{aligned}
  \end{rcases*}
\end{align}
This system of equations must be supplemented with boundary conditions at
the channel edges. By generalization of the DS boundary conditions,
Eq.~\eqref{eq:DS-BCs}, we impose
\begin{subequations}
\begin{align}
  \label{eq:boundary_conditions}
  \begin{rcases*}
    \begin{aligned}
      \rho &= \rho_0\quad \text{at }x=0,
      \\
      J_x&=v_x\,\rho = v_{x0}\,\rho_0\quad \text{at }x=L,
    \end{aligned}
  \end{rcases*}
\end{align}
where $\vec J=(J_x, J_y)$ is the flux vector. Here, we require that the
density $\rho$ and normal flux $J_x$ be held fixed at the left and right
boundary, respectively, similarly to the DS
setting~\cite{DyakonovShur1993}.  Furthermore, we need a
boundary condition for the tangential (or, lateral) flux, $J_y=v_y\rho$. 

We study the following two scenarios. In {\em Case~I},
$J_y$ takes mutually opposite values at the boundaries, viz.,
\begin{align}\label{eq:Jy-bc-opposite-val}
J_y\big|_{x=0}=-J_y\big|_{x=L}.	
\end{align}
Alternatively, in {\em Case II}, $J_y$ vanishes at the right
boundary, viz.,
\begin{align}\label{eq:Jy-bc-zero-right}
J_y=0\quad\mbox{at }x=L.
\end{align}
\end{subequations}
 The imposition of either Eq.~\eqref{eq:Jy-bc-opposite-val} or
Eq.~\eqref{eq:Jy-bc-zero-right} provides a third (scalar) boundary
condition in the model. 

Condition~\eqref{eq:Jy-bc-opposite-val} is deemed
as physically reasonable, since it is consistent with the property that
the $\vec B$ field tends to generate a centripetal force on an electron along
a circle of diameter $L$ centered at $x=L/2$.  Indeed, Case~I yields a steady state with $J_y=\pm eB L\rho_0/(2m_e)$ at the two ends of the channel (Sec.~\ref{sssec:exact-ss}). By our linearized-stability formulation around this steady state, the resulting perturbations of these boundary values come from the related eigenfunctions (eigenmodes) and are not uniquely defined. In an actual physical setting, these values should be determined uniquely at times $t>0$ via the prescription of suitable initial data (at $t=0$) for the time-dependent hydrodynamic equations. This task is distinct from our spectral problem. For a discussion, see Sec.~\ref{subsec:discussion}.   On the other hand,
condition~\eqref{eq:Jy-bc-zero-right} of Case~II is imposed for comparison purposes.
The model of Eqs.~\eqref{eq:shallow_water_model_with_B}
and~\eqref{eq:boundary_conditions} with Eq.~\eqref{eq:Jy-bc-opposite-val}
or~\eqref{eq:Jy-bc-zero-right}  can   be complemented with initial
conditions for $(\rho, J_x, J_y)$, which we omit here.

A few more remarks are in order. First, the governing differential
equations are not invariant under time reversal (by $t\mapsto -t$) if
$B\neq 0$. Second, the particular choice of Eq.~\eqref{eq:Jy-bc-zero-right}
yields a system of boundary conditions that closely imitates the DS
setting. In this case, the flux vector $\vec J$ near the right boundary
tends to locally reduce to the form $(\text{const}, 0)$, which resembles
the 1D flow of the DS model.   This choice is mathematically
convenient. For example, with Eq.~\eqref{eq:Jy-bc-zero-right} the fluid
spectrum \emph{smoothly} reduces to that of the DS model in the limit $\vec
B\to 0$. In contrast, this limit exhibits an anomaly under
condition~\eqref{eq:Jy-bc-opposite-val}, as we discuss later on (Sec.~\ref{subsec:spectrum}). Third, for each choice of the boundary condition for $J_y$,
the electronic steady state has a nontrivial dependence on $x$ if $B\neq
0$.   This feature contrasts the constant steady state
in~\cite{DyakonovShur1993}, and dramatically affects the fluid
spectrum.

Notice that we have applied the usual convention of particles with a
positive charge $e$ (i.\,e., $e>0$). The case with a
negative charge (if $e\mapsto  -e$) leads to dynamics identical to those
with fixed positive charge $e$  and flipped sign of $B$ ($B\mapsto
-B$). In this latter setting, the mirror symmetry of the governing
equations entails that $(\rho, v_x, v_y)\mapsto  (\rho, v_x, -v_y)$ with
$y\mapsto  -y$. In other words, if $(\rho, v_x, v_y)$ is a solution of
the model for given $B$ and fixed $e$ then $(\rho,
v_x, -v_y)$  must also be a solution if $(y,B)$ is replaced by $-(y,B)$.
Thus, it suffices to assume that $B>0$ throughout. To simplify the
analysis, we additionally enforce \emph{translation invariance} and
restrict our attention to $y$-independent solutions, setting
\begin{equation}\label{eq:y-indep}
  \frac{\partial}{\partial y}(\rho, J_x, J_y)=0\quad \mbox{everywhere}.
\end{equation}
Despite this restriction, the tangential velocity, $v_y$, is nonzero in the
channel ($0<x<L$) if $B\neq 0$.

\subsection{Rescaled variables}
\label{subsec:rescaling}

Next, we introduce rescaled variables and nondimensionalize the governing
equations, for later numerical and algebraic convenience. Recall the
quantities $\rho_0$ and $v_{x0}$, which are the reference values for the
electronic number density and normal velocity from the boundary
conditions in Eq.~\eqref{eq:boundary_conditions}~\cite{DyakonovShur1993}.

First, we introduce a reference magnetic field, $B_0>0$. The respective
dimensionless parameter $\beta$ is defined by
\begin{equation*}
  \beta= B/B_0>0.
\end{equation*}
Recall the sound speed, $s_0=\sqrt{a\rho_0}$ (Sec.~\ref{subsec:DS-stabil}).

We rescale the dependent variables by setting
\begin{align*}
  \breve{v}_x=\frac{v_x}{v_{x0}},\quad
  \breve{v}_y=\frac{v_y}{s_0},\quad
  \breve{\rho}=\frac{\rho}{\rho_0}. \quad
\end{align*}
The rescaled $x$ coordinate and channel width are
\begin{equation*}
  \breve{x} = \frac{x}{\ell},\quad
  \breve{l}=\frac{L}{\ell}\quad\mbox{where}\;\ell=\frac{s_0m_e}{e B_0};
\end{equation*}
thus, $0\le \breve{x}\le \breve{l}$. Notably, the hydrodynamic
  quantities $(\breve v_x, \breve v_y, \breve \rho)$ are scaled
by $v_{x0}$, $s_0$ and $\rho_0$ which are independent of the parameter
$\beta$  and thus of the magnetic field $B$. This ensures that
the rescaling remains well defined in the limit $\beta\to0$.
We will fix the reference field $B_0$ by setting $\breve l=1$ for our numerics 
(Sec.~\ref{sec:numerical_results}). This value entails $B_0=\tfrac{s_0
m_e}{e L}$ which  yields a conversion formula for the frequencies of
the fluid spectrum, as discussed in Sec.~\ref{subsec:discussion}. For the
sake of generality, we use an arbitrary yet $\beta$-independent $\breve l$ ($\breve l>0$) in the
present section and in Sec.~\ref{sec:linear_stability}.  

The equations of motion in the scaled variables read
\begin{align}
  \label{eq:scaled-model_with_B}
  \begin{rcases*}
    \begin{aligned}
      \left(\frac{\partial}{\partial \breve t} +
      \breve v_x\frac{\partial}{\partial \breve x} \right)\breve v_x
      \;&=\;
      M^{-2} \left(-\frac{\partial\breve \rho}{\partial \breve x} +
      \beta\,\breve v_y\right),
      \\
      \left(\frac{\partial}{\partial \breve t} +
      \breve v_x\frac{\partial}{\partial \breve x}\right)\breve v_y
      \;&=\;
      -  \beta\,\breve v_x,
      \\
      \frac{\partial \breve \rho}{\partial \breve t} +
      \frac{\partial}{\partial \breve x}(\breve \rho\,\breve v_x)\;&=\; 0,
    \end{aligned}
  \end{rcases*}
\end{align}
where (for $M=v_{x0}/s_0$)
\begin{equation*}
  \breve t=\frac{v_{x0}}{\ell}t=M\frac{eB_0}{m_e}t.
\end{equation*}
  In view of conditions~\eqref{eq:Jy-bc-opposite-val}
and~\eqref{eq:Jy-bc-zero-right}, Eq.~\eqref{eq:scaled-model_with_B} is
subject to one of the following two sets of boundary conditions.
Case I corresponds to
\begin{align}
  \label{eq:scaled-BCs-opposite}
  \begin{rcases*}
    \begin{aligned}
      \breve\rho &= 1\quad \text{at }\breve x=0,
      \\
      \breve J_x&=\breve v_x\,\breve \rho = 1 \quad \text{at }\breve x=\breve l,\\
      &\breve J_y\big|_{x=0}+\breve J_y\big|_{x=\breve l}=0.
    \end{aligned}
  \end{rcases*}
\end{align}
Case II amounts to
\begin{align}
  \label{eq:scaled-BCs}
  \begin{rcases*}
    \begin{aligned}
      \breve\rho &= 1\quad \text{at }\breve x=0,
      \\
      \breve J_x&=\breve v_x\,\breve \rho = 1,\quad
      \breve J_y=\breve v_y \breve \rho=0 \quad \text{at }\breve x=\breve l.
    \end{aligned}
  \end{rcases*}
\end{align}

\subsection{Steady-state solution for $B\neq 0$}
\label{subsec:steady_state}

 The magnetically driven system of
Eq.~\eqref{eq:scaled-model_with_B} together with
Eq.~\eqref{eq:scaled-BCs-opposite} or~\eqref{eq:scaled-BCs}  admits a
time-independent solution,   which we will describe in the
rescaled variables $(\breve \rho, \breve v_x, \breve v_y)$  for
each choice of the boundary condition for $\breve J_y$.   For
details on the analytical derivation, see Appendix~\ref{app:steady_state}.

\subsubsection{Exact steady state}
\label{sssec:exact-ss}

Let $( \breve{\rho}^s, \breve{v}_x^s, \breve{v}_y^s)$ denote the
steady-state solution.   We distinguish two cases.

\paragraph*{Case I: Mutually opposite boundary values of lateral flux.} Let us
consider a time-independent solution to Eqs.~\eqref{eq:scaled-model_with_B}
and~\eqref{eq:scaled-BCs-opposite}. After some algebra, this solution is
cast into the following algebraic form (for $0\le\breve x\le\breve l$):
\begin{gather}
  \breve{v}_x^s(\breve x) = \big(\breve \rho^s(\breve x)\big)^{-1},
  \quad
  \breve{v}_y^s(\breve x) = \beta\,\left(\tfrac{\breve{l}}{2} - \breve{x}\right),
  \notag\\
  \breve{\rho}^s(\breve x) = b^{1/3} \,\epsilon(\breve x)^{-2/3}\,
  \big\{\tfrac23\cos(\tfrac{\theta(\breve x)}{3})+\tfrac13\big\}.
  \label{eq:rescaled-ss}
\end{gather}
Here, we set $b=\tfrac{1}{2}M^2 = \tfrac{1}{2}\frac{v_{x0}^2}{s_0^2}>0$,
\begin{equation}
  \label{eq:epsilon-x}
  \epsilon(\breve x)= \frac{\sqrt{b}}{\big[1 + b + \tfrac{\beta^2}{2}\breve
  x(\breve{l} - \breve{x})\big]^{3/2}},
\end{equation}
and by using a suitable analytic continuation of the inverse cosine
(Appendix~\ref{app:steady_state}), we introduce the functions
\begin{equation}
  \theta(\breve{x})=
  \begin{cases}
    \begin{aligned}
      \label{eq:theta-ss}
      \cos^{-1}\big(1-\tfrac{27}{2}\epsilon(\breve x)^2\big),\quad
      &\text{for }b<\tfrac{1}{2},
      \\[0.2em]
      2\pi\,-\,\cos^{-1}\big(1-\tfrac{27}{2}\epsilon(\breve x)^2\big),\quad
      &\text{for }b>\tfrac{1}{2}.
    \end{aligned}
  \end{cases}
\end{equation}
In the above, the function $\cos^{-1}(\cdot)$ takes (real) values only
in its principal branch ($0\le \cos^{-1}(\cdot)< \pi$). The
cases with $b<1/2$ and $b>1/2$ amount to $v_{x0}<s_0$ and $v_{x0}> s_0$,
respectively. We still refer to these cases as the
subsonic and supersonic regime, respectively. Evidently, by
Eqs.~\eqref{eq:rescaled-ss}--\eqref{eq:theta-ss}, the variables
$\breve{\rho}^s$ and $\breve{v}_x^s$ are symmetric (i.e., even) with
respect to the channel center, $\breve x=\breve l/2$; whereas
$\breve{v}_y^s$ is anti-symmetric (odd).   Notice that the sign of $\breve
v_y^s$ is reversed but $\breve v_x^s$ and $\breve\rho^s$ remain invariant
if the sign of either $B$ or $e$ is switched, as expected by the mirror
symmetry of the governing equations. If $B\to 0$, this solution
continuously reduces to that of the DS model~\cite{DyakonovShur1993}; for
example, notice that $\breve v_y\to 0$ as $B\to 0$.

\paragraph*{Case II: Zero lateral flux at the right boundary.}
We seek a time-independent solution to
Eqs.~\eqref{eq:scaled-model_with_B} and~\eqref{eq:scaled-BCs}. By analogy
with Eqs.~\eqref{eq:rescaled-ss} and~\eqref{eq:epsilon-x}, we obtain  

\begin{gather}
  \breve{v}_x^s(\breve x) = \big(\breve \rho^s(\breve x)\big)^{-1},
  \quad
  \breve{v}_y^s(\breve x) = \beta\,\big(\breve{l} - \breve{x}\big),
  \notag\\
  \breve{\rho}^s(\breve x) = b^{1/3} \,\epsilon(\breve x)^{-2/3}\,
  \big\{\tfrac23\cos(\tfrac{\theta(\breve x)}{3})+\tfrac13\big\},
  \label{eq:rescaled-ss-alt}
\end{gather}
and 
\begin{equation}
  \label{eq:epsilon-x-alt}
  \epsilon(\breve x)= \frac{\sqrt{b}}{\big[1 + b + \tfrac{\beta^2}{2}\breve
  x(2\,\breve{l} - \breve{x})\big]^{3/2}},
\end{equation}
 where $\theta(\breve x)$ is defined by Eq.~\eqref{eq:theta-ss}.
Note that in this case the tangential velocity, $\breve{v}_y^s$, results
from its counterpart in Eq.~\eqref{eq:rescaled-ss} by a mere shift of
$\breve x$ by $\breve l/2$, i.e., $\breve x\mapsto \breve x-\breve l/2$. On
the other hand, the $\breve x$-dependence of $\epsilon(\breve x)$ comes
from the appropriate integration of $\breve{v}_y^s(\breve x)$ (see
Appendix~\ref{app:steady_state}).  Again, as $B\to 0$, this solution
continuously reduces to that of the DS model~\cite{DyakonovShur1993}.  

An interesting, yet not surprising, mathematical feature of the above
steady states should be pointed out.   The distinct effect of
each kinetic (subsonic or supersonic) regime can be connected to a
particular branch of an intrinsically multivalued function, namely, the
generalized inverse cosine $\cos^{-1}z$,   as is indicated by the
definition of $\theta(\breve x)$;   see
Appendix~\ref{app:steady_state}. This \emph{branching} is manifested in the
steady-state normal velocity and density; in contrast, the lateral velocity
does not distinguish between the two regimes. For fixed $\beta$ and $\breve
x$  ($\breve x\neq 0,\,\breve l$),   we see that the
steady-state density and normal flux as a function of $b$ exhibit a jump at
the transition point, $b=1/2$. This structure of the steady state endows
the magnetic fluid system with linear stability and instability spectra
qualitatively different from those of the DS case ($B=0$), as discussed in
Sec.~\ref{sec:numerical_results}.  We should mention that,
for Case~I, in the limit
$\beta\to 0$ ($B\to 0$) the fluid spectrum  exactly reduces to the union of the spectrum of the DS model and a spectrum of geometric character with a nonzero lateral flux in the channel. This limit   is discussed in
Secs.~\ref{subsec:discussion} and~\ref{sec:conclusion}.

\subsubsection{Asymptotics for weak magnetic field, $|\beta|\breve l\ll 1$}
\label{sssec:asymptotics-ss}

 Formulas~\eqref{eq:rescaled-ss} and~\eqref{eq:rescaled-ss-alt}
  are  simplified for a weak magnetic field (see
Appendix~\ref{app:steady_state}). By inspection of
 Eqs.~\eqref{eq:epsilon-x} and~\eqref{eq:epsilon-x-alt},
  we realize that the relevant parameter is
\begin{equation*}
  \beta\breve l=\frac{eB L}{s_0 m_e},
\end{equation*}
which expresses the strength of the magnetic Lorentz force with magnitude
$e s_0 |B|$ relative to the mechanical (centripetal-like) intrinsic force
of magnitude $m_e s_0^2/L$ on an electron. The assumption for a weak
magnetic field means that the latter force dominates, i.e., $|\beta\breve
l|\ll 1$. Consider $\beta$ as real (positive or negative) in this section.

It turns out that the approximation for $\breve\rho^s$ also depends on the
proximity of the kinetics to the transition point, $b=1/2$. In fact, for
$|\beta|\breve l\ll 1$ and arbitrary positive $b$, we  distinguish the
cases with $|\beta| \breve l\ll \sqrt{|1-2b|}$ and $\beta \breve l=\mathcal
O\left(\sqrt{|1-2b|} \right)$. In the former case, we can simplify $\breve
\rho^s$ by regular perturbations, ignoring the transition point. In the
latter case, $\breve\rho^s$ is amenable to singular perturbations since the
kinetics of the transition point, if $b$ is near $1/2$, significantly
interfere with perturbations in $\beta$.

Consider $\beta \breve l\ll \sqrt{|1-2b|}$, which signifies an
\emph{extremely weak} magnetic field.   Let us discuss the
scenario of Case~I. By
Eq.~\eqref{eq:rescaled-ss},   we calculate (see
Appendix~\ref{app:steady_state})
\begin{subequations}
\label{eqs:scaled-rho-ss-smallB}
\begin{equation}
  \label{eq:scaled-rho-ss-smallB-reg}
  \breve\rho^s(\breve x) \simeq 1+\frac{\beta^2}{2(1-2b)} \breve x (\breve
  l-\breve x),  \quad 0\le \breve x\le \breve l,
\end{equation}
which is recognized as a regular perturbation of the DS steady state.
Notice the absence of a linear-in-$\beta$ term in the above expansion, as
expected because the sign reversal of $B$ leaves $\breve\rho^s$
invariant. Similarly, the rescaled normal velocity, $\breve
v_x^s=(\breve\rho^s)^{-1}$, exhibits an $\mathcal O(\beta^2\breve l^2)$
correction to its DS steady-state value, for $\beta=0$. On the other hand, we
have  $\breve v_y^s=\beta(\breve l/2-\breve x)$.  

Next, consider  $\beta \breve l=\mathcal O\left(\sqrt{|1-2b|} \right)$,
which provides a critical scaling for $\beta$ near the transition point.
 By Eq.~\eqref{eq:rescaled-ss},   we obtain (see
Appendix~\ref{app:steady_state})
\begin{equation}
  \label{eq:scaled-rho-ss-smallB-sing}
  \breve\rho^s(\breve x) \simeq
  1+\frac{|\beta|}{\sqrt{3}}\,\mathrm{sgn}(1-2b)\,  \sqrt{\breve x(\breve
  l-\breve x)},  
\end{equation}
\end{subequations}
for $0\le \breve x\le \breve l$. In the above, $\mathrm{sgn}$ denotes the
signum function; $\mathrm{sgn}(x)=1$ if $x>0$ and $\mathrm{sgn}(x)=-1$ if
$x<0$. Asymptotic formulas~\eqref{eq:scaled-rho-ss-smallB-reg}
and~\eqref{eq:scaled-rho-ss-smallB-sing} are distinct and non overlapping.
These expressions are connected only through the more complicated
Eqs.~\eqref{eq:rescaled-ss} and~\eqref{eq:theta-ss}.

 The scenario of Case~II, which leads to Eq.~\eqref{eq:rescaled-ss-alt}, can be studied
similarly. Formulas~\eqref{eq:scaled-rho-ss-smallB-reg}
and~\eqref{eq:scaled-rho-ss-smallB-sing} are applicable with the factor
$\breve x (\breve l-\breve x)$ replaced by $\breve x (2\breve l-\breve x)$.


\section{Linear stability formulation}
\label{sec:linear_stability}

In this section, by linear stability theory we formulate an eigenvalue
problem for the model based on  
Eqs.~\eqref{eq:scaled-model_with_B} and~\eqref{eq:scaled-BCs-opposite}, or
Eqs.~\eqref{eq:scaled-model_with_B} and~\eqref{eq:scaled-BCs}, with $\beta\neq
0$.   We derive a system of linear equations for the
hydrodynamic variables, and discuss its properties for a sufficiently weak
magnetic field. For more details, see
Appendix~\ref{app:linearized_stability}.

In the spirit of Sec.~\ref{subsec:DS-stabil} for the DS model, let us
consider the scaled (dimensionless) flux vector $(\breve J_x, \breve
J_y)=(\breve\rho\breve v_x, \breve \rho \breve v_y)$ and  density $\breve
\rho$, and perturb them around a steady-state solution of
Sec.~\ref{subsec:steady_state}. Hence, we write
\begin{equation*}
  (\breve J_x, \breve J_y)=(\breve
  J_x^s(\breve x), \breve J_y^s(\breve x))+(\breve\jmath_1(\breve x),
  \breve\jmath_2(\breve x))e^{-\breve \Lambda \breve t},
\end{equation*}
where $(\breve J_x^s, \breve J_y^s)=(\breve\rho^s\breve v_x^s, \breve
\rho^s \breve v_y^s)$, while
\begin{equation*}
  \breve \rho=\breve \rho^s(\breve x)+\breve \rho_1(\breve x)e^{-\breve
  \Lambda \breve t},\quad 0<\breve x< \breve l.
\end{equation*}
In the above, the physically admissible values of the parameter
$\breve\Lambda$ must be determined consistently with the governing
equations of motion and boundary conditions for the perturbation variables
$\breve \rho_1$, $\breve\jmath_1$, $\breve\jmath_2$. Note that $\breve
\Lambda$ is dimensionless; its dimensional counterpart, $\Lambda$, has
units of inverse time (or, frequency) and is related to $\breve \Lambda$ by
\begin{align*}
  \Lambda\;=\; \frac{eB_0v_{x0}}{m_e s_0}\,\breve \Lambda =
  \frac{eB_0\sqrt{2b}}{m_e}\,\breve \Lambda.
\end{align*}
Recall that in the rescaled variables the time is expressed via $\breve t$
in units of $\frac{m_e}{eB_0} \frac{s_0}{v_{x0}}$  which equals
$\tfrac{L}{v_{x0}}$ if $\breve l=1$.   Since $\Lambda=\im
\omega$, we write $\breve \Lambda=\im
\breve \omega$ with the rescaled frequency $\breve \omega=\frac{m_e s_0}{e
B_0 v_{x0}}\omega$   which becomes $\breve
\omega=\tfrac{L}{v_{x0}}\omega$ for $\breve l=1$.

\subsection{Governing motion laws}
\label{subsec:stabil-magn-govern-laws}
The formulation with rescaled variables is summarized as follows. First,
 the systems of Eqs.~\eqref{eq:scaled-model_with_B}
and~\eqref{eq:scaled-BCs-opposite}, and Eqs.~\eqref{eq:scaled-model_with_B}
and~\eqref{eq:scaled-BCs}   are converted to systems of
nonlinear equations for the vector-valued function $(\breve J_x, \breve
J_y, \breve \rho)$, as shown in Appendix~\ref{app:linearized_stability}. By
linearizing each system of governing equations around the corresponding
steady state, we obtain in matrix form the related eigenvalue problem for
$\breve\Lambda$ and the vector-valued function $\overline{U}(\breve x) =
(\breve \jmath_1(\breve x), \breve\jmath_2(\breve x), \breve \rho_1(\breve
x))^T$ where the subscript $T$ denotes the transpose of a vector, viz.,
\begin{gather}
  \label{eq:eigenvalue_problem}
  \frac{\partial}{\partial \breve x}\Big\{\mathbb{A}(\breve x)
  \overline U(\breve x) \Big\} \;+\; \beta \mathbb{B}\,\overline
  U(\breve x) \;=\; \im\, \breve\omega\, \overline U(\breve x),
\end{gather}
where
\begin{gather*}
  \mathbb{A}(\breve x)=
  \begin{pmatrix}
    2\breve{v}_x^s & 0 & \frac{\breve{\rho}^s}{2b} - (\breve v_x^s)^2
    \\[0.5em]
    \breve{v}_y^s & \breve{v}_x^s & - \breve{v}_x^s \breve{v}_y^s
    \\[0.5em]
    1 & 0 & 0
  \end{pmatrix},\;
  \mathbb{B}=
  \begin{pmatrix}
    0 & - \frac{1}{2b} & 0 \\
    1 & 0 & 0 \\
    0 & 0 & 0
  \end{pmatrix},
\end{gather*}
for $0< \breve x < \breve{l}$. Recall that the functions $\breve \rho^s$,
$\breve v_x^s$ and $\breve v_y^s$ depend on $\beta$. We also apply the
following boundary conditions:  
\begin{gather}
  \label{eq:eigenvalue_problem_boundary_conditions}
  \text{Case I}:\   \breve \rho_1(0) = 0,\ \breve\jmath_1(\breve l) = 0,
  \ \breve \jmath_2(0)=-\breve\jmath_2(\breve l);
\end{gather}
or, alternatively,
\begin{gather}
  \label{eq:eigenvalue_problem_boundary_conditions-alt}
  \text{Case II}:\ \breve \rho_1(0) = 0,\ \breve\jmath_1(\breve l) = 0,
  \ \breve\jmath_2(\breve l) = 0.
\end{gather}
Equation~\eqref{eq:eigenvalue_problem} together with
Eq.~\eqref{eq:eigenvalue_problem_boundary_conditions}
or~\eqref{eq:eigenvalue_problem_boundary_conditions-alt} should yield the
desired sets of eigen-frequencies, $\breve\omega$.   We have
been unable to solve these eigenvalue problems analytically, because of the
complicated dependence of $\mathbb{A}(\breve x)$ on $\breve x$ via the
steady-state solution. In Sec.~\ref{sec:numerical_results}, we present
numerical computations for the stability and instability spectra,  and compare related features of Cases~I and~II.

\subsection{Scaling law for extremely weak magnetic field}
\label{subsec:stabil-magn-weakB}

Next, we outline how the deviation of the rescaled eigen-frequency $\breve
\omega$ for $\beta\neq 0$ from its corresponding unperturbed value $\breve
\omega^0$ of the DS model scales with $\beta$ if
$(\beta\breve l)^2\ll |1-2b|$. We invoke results of
Sec.~\ref{sssec:asymptotics-ss} for the expansion of the steady state.
  Our discussion here concerns both eigenvalue problems of
Sec.~\ref{subsec:stabil-magn-govern-laws}. For small $|\beta|$, we expand
the solution of the linearized model around the DS solution  for the fields
and spectrum.  

We use the fact that the steady state solution has an expansion in integer
powers of $\beta$ in this asymptotic regime. Consequently, the matrix
$\mathbb{A}(\breve x)$ of  Eq.~\eqref{eq:eigenvalue_problem} admits a
similar expansion in powers of $\beta$. We apply regular perturbation
theory to the ensuing system of equations. Hence, in
Eq.~\eqref{eq:eigenvalue_problem} we try the formal expansions
\begin{equation*}
  \overline{U} \simeq \overline{U}_0+\beta^c\, \overline{\mathfrak U},\quad
  \breve \omega\simeq \breve \omega^0+\beta^\nu\, \breve \omega^{dev}.
\end{equation*}
Here, $(\overline{U}_0, \breve \omega^0)$ represents the rescaled solution
of the  DS model,   the unknown
coefficients $\overline{\mathfrak U}$ and $\breve \omega^{dev}$ are assumed
independent of $\beta$, and the exponents $c$ and $\nu$ must be determined
consistently with the governing equations. In this procedure, we take into
account that $\overline{U}_0$ has zero lateral flux  
($\breve\jmath_{2,0}=0$ everywhere),   and the matrix
$\mathbb{B}$ is independent of $\beta$ while $\mathbb{A}=\mathbb{A}_0+\beta
\mathbb{A}_1+\ldots$ where $\mathbb{A}_1$ is sparse with zero first and
third rows. After some algebra, dominant-balance arguments yield $c=1$ and
$\nu=2$, with $\overline{\mathfrak U}$ containing only one nonzero
component  in the interior of the channel,   namely,
the lateral flux. These findings are compatible with the mirror symmetry of
the governing motion laws (when $\beta\mapsto -\beta$).

Hence, we reach the conclusion that
\begin{equation}
  \label{eq:beta-scaling-eigenf}
  \breve \omega-\breve \omega^0=\mathcal O(\beta^2\breve{l}^2)\quad
  \mbox{if}\ \ (\beta\breve l)^2\ll |1-2b|,
\end{equation}
 for each case of the boundary condition for the tangential flux,
where $\breve \omega^0$ corresponds to the DS spectrum.   This
scaling prediction is corroborated by numerical simulations
(Sec.~\ref{sec:numerical_results}). We expect that this prediction is
modified in the regime where $(\beta\breve l)^2=\mathcal O(|1-2b|)$. This
case requires the application of singular perturbations to the governing
equations, and is not treated here.

%
\section{Computed spectra and gap}
\label{sec:numerical_results}
In this section, we numerically solve the eigenvalue problem
 described by Eq.~\eqref{eq:eigenvalue_problem} with the boundary
conditions of Eq.~\eqref{eq:eigenvalue_problem_boundary_conditions}
or~\eqref{eq:eigenvalue_problem_boundary_conditions-alt}, for Case~I or~II, respectively.
 \ 
For this purpose, we discretize the related differential operator with a
stabilized finite difference scheme. Our numerical method achieves a
well-controlled discretization error. For details on the procedure, see
Appendix~\ref{app:numerical_approach}.

First, we provide numerical results for the instability and stability
spectra, i.e., the complex eigen-frequencies $\breve\omega$, for different
values of the parameters $\beta$ and $\delta=b-1/2$. The former parameter
expresses the strength of the magnetic field and the latter measures the
deviation from the transition point.  For our numerical
computations, we have set $\breve l=1$ throughout. \ 
Second, we compute the magnetically controlled spectral gap, showing numerically how
this varies with $\beta$ and $\delta$. Thirdly, we discuss implications of
our results.

\subsection{Discrete frequency spectrum}
\label{subsec:spectrum}
\begin{figure}[!]
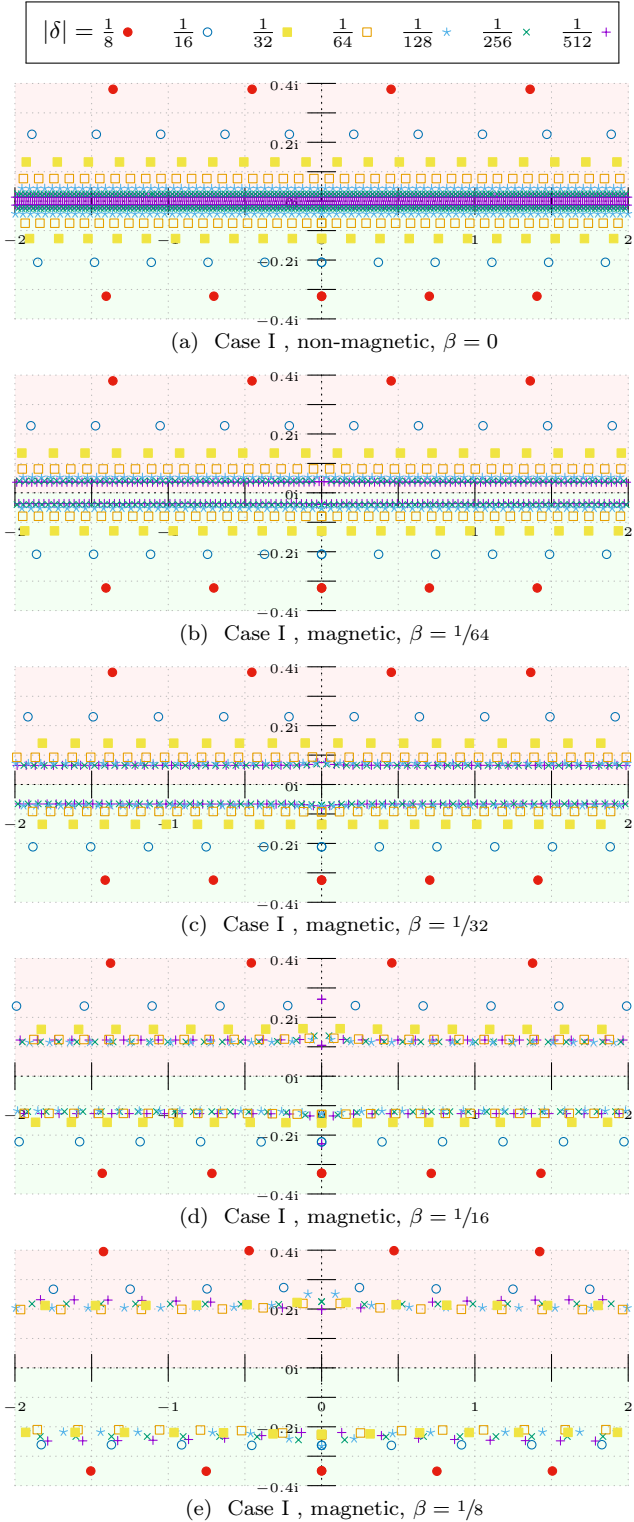

  \centering
  \vspace{0.75em}
  \fbox{\begin{tikzpicture}[gnuplot]
\tikzset{every node/.append style={font={}}}
\gpsetlinetype{gp lt axes}
\gpsetdashtype{gp dt axes}
\gpsetlinewidth{1.00}
\gpsetpointsize{4.00}
\gpcolor{color=gp lt color border}
\node[gp node right] at (2.600,0.334) {$|\delta|=\tfrac{1}{8}$};
\gpcolor{rgb color={0.898,0.118,0.063}}
\gp3point{gp mark 7}{}{(2.640,0.334)}
\gpcolor{color=gp lt color border}
\node[gp node right] at (3.660,0.334) {$\tfrac{1}{16}$};
\gpcolor{rgb color={0.000,0.447,0.698}}
\gp3point{gp mark 6}{}{(3.700,0.334)}
\gpcolor{color=gp lt color border}
\node[gp node right] at (4.720,0.334) {$\tfrac{1}{32}$};
\gpcolor{rgb color={0.941,0.894,0.259}}
\gp3point{gp mark 5}{}{(4.760,0.334)}
\gpcolor{color=gp lt color border}
\node[gp node right] at (5.780,0.334) {$\tfrac{1}{64}$};
\gpcolor{rgb color={0.902,0.624,0.000}}
\gp3point{gp mark 4}{}{(5.820,0.334)}
\gpcolor{color=gp lt color border}
\node[gp node right] at (6.840,0.334) {$\tfrac{1}{128}$};
\gpcolor{rgb color={0.337,0.706,0.914}}
\gp3point{gp mark 3}{}{(6.880,0.334)}
\gpcolor{color=gp lt color border}
\node[gp node right] at (7.900,0.334) {$\tfrac{1}{256}$};
\gpcolor{rgb color={0.000,0.620,0.451}}
\gp3point{gp mark 2}{}{(7.940,0.334)}
\gpcolor{color=gp lt color border}
\node[gp node right] at (8.960,0.334) {$\tfrac{1}{512}$};
\gpcolor{rgb color={0.580,0.000,0.827}}
\gp3point{gp mark 1}{}{(9.000,0.334)}
\gpdefrectangularnode{gp plot 1}{\pgfpoint{0.460cm}{0.676cm}}{\pgfpoint{13.417cm}{3.501cm}}
\end{tikzpicture}}\\[-0.25em]
  \subfloat[ Case I , non-magnetic, $\beta=0$]{\parbox{0.5\textwidth}{\hspace{-0.75em}\input{nbc-beta-0.tex}\\[-1.1em]}}\\[-0.25em]
  \subfloat[ Case I , magnetic, $\beta=\sfrac{1}{64}$]{\parbox{0.5\textwidth}{\hspace{-0.75em}\input{nbc-beta-164.tex}\\[-1.1em]}}\\[-0.25em]
  \subfloat[ Case I , magnetic, $\beta=\sfrac{1}{32}$]{\parbox{0.5\textwidth}{\hspace{-0.75em}\input{nbc-beta-132.tex}\\[-1.1em]}}\\[-0.25em]
  \subfloat[ Case I , magnetic, $\beta=\sfrac{1}{16}$]{\parbox{0.5\textwidth}{\hspace{-0.75em}\input{nbc-beta-116.tex}\\[-1.1em]}}\\[-0.25em]
  \subfloat[ Case I , magnetic, $\beta=\sfrac{1}{8}$]{\parbox{0.5\textwidth}{\hspace{-0.75em}\input{nbc-beta-18.tex}\\[-1.1em]}}\\[0.2em]
  \caption{
     Case I:\ 
    Computed eigen-frequencies $\breve\omega_n$ from the eigenvalue problem
    of Eqs.~\eqref{eq:eigenvalue_problem}
and~\eqref{eq:eigenvalue_problem_boundary_conditions}. The magnetic-field related values are $\beta=0$,
    $\sfrac{1}{64}$, $\sfrac{1}{32}$, $\sfrac{1}{16}$, $\sfrac{1}{8}$; and the 
    kinetic-parameter values  are $\delta=\pm\sfrac{1}{8}$,
    $\pm\sfrac{1}{16}$, $\pm\sfrac{1}{32}$, $\pm\sfrac{1}{64}$,
    $\pm\sfrac{1}{128}$, $\pm\sfrac{1}{256}$, $\pm\sfrac{1}{512}$. The
    green (red) regions contain eigen-frequencies of the stability (instability) spectrum, for $\delta>0$ ($\delta<0$).
  }
  \label{fig:spectrum}
\end{figure}
\begin{figure}[!]
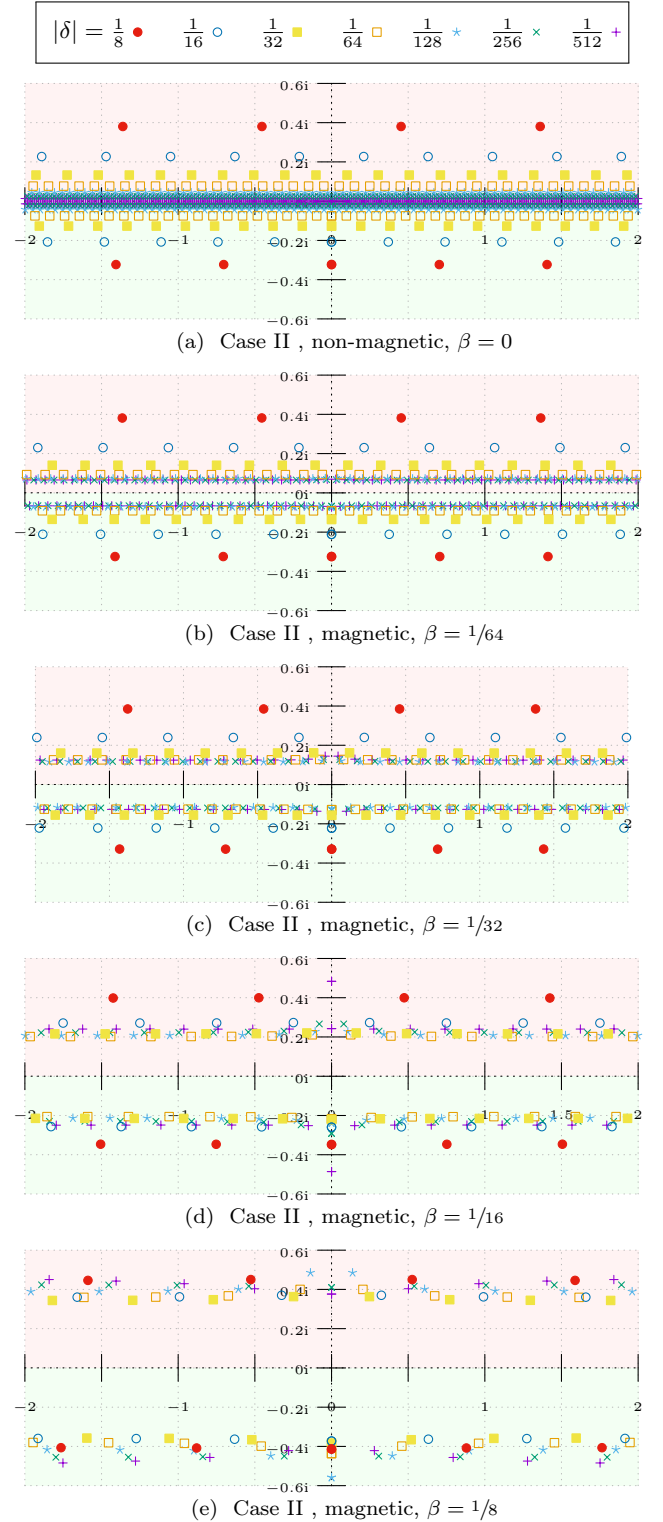

  \centering
  \vspace{0.75em}
  \fbox{\begin{tikzpicture}[gnuplot]
\tikzset{every node/.append style={font={}}}
\gpsetlinetype{gp lt axes}
\gpsetdashtype{gp dt axes}
\gpsetlinewidth{1.00}
\gpsetpointsize{4.00}
\gpcolor{color=gp lt color border}
\node[gp node right] at (2.600,0.334) {$|\delta|=\tfrac{1}{8}$};
\gpcolor{rgb color={0.898,0.118,0.063}}
\gp3point{gp mark 7}{}{(2.640,0.334)}
\gpcolor{color=gp lt color border}
\node[gp node right] at (3.660,0.334) {$\tfrac{1}{16}$};
\gpcolor{rgb color={0.000,0.447,0.698}}
\gp3point{gp mark 6}{}{(3.700,0.334)}
\gpcolor{color=gp lt color border}
\node[gp node right] at (4.720,0.334) {$\tfrac{1}{32}$};
\gpcolor{rgb color={0.941,0.894,0.259}}
\gp3point{gp mark 5}{}{(4.760,0.334)}
\gpcolor{color=gp lt color border}
\node[gp node right] at (5.780,0.334) {$\tfrac{1}{64}$};
\gpcolor{rgb color={0.902,0.624,0.000}}
\gp3point{gp mark 4}{}{(5.820,0.334)}
\gpcolor{color=gp lt color border}
\node[gp node right] at (6.840,0.334) {$\tfrac{1}{128}$};
\gpcolor{rgb color={0.337,0.706,0.914}}
\gp3point{gp mark 3}{}{(6.880,0.334)}
\gpcolor{color=gp lt color border}
\node[gp node right] at (7.900,0.334) {$\tfrac{1}{256}$};
\gpcolor{rgb color={0.000,0.620,0.451}}
\gp3point{gp mark 2}{}{(7.940,0.334)}
\gpcolor{color=gp lt color border}
\node[gp node right] at (8.960,0.334) {$\tfrac{1}{512}$};
\gpcolor{rgb color={0.580,0.000,0.827}}
\gp3point{gp mark 1}{}{(9.000,0.334)}
\gpdefrectangularnode{gp plot 1}{\pgfpoint{0.460cm}{0.676cm}}{\pgfpoint{13.417cm}{3.501cm}}
\end{tikzpicture}}\\[-0.25em]
  \subfloat[ Case II , non-magnetic, $\beta=0$]{\parbox{0.5\textwidth}{\hspace{-0.75em}\input{beta-0.tex}\\[-1.1em]}}\\[-0.25em]
  \subfloat[ Case II , magnetic, $\beta=\sfrac{1}{64}$]{\parbox{0.5\textwidth}{\hspace{-0.75em}\input{beta-164.tex}\\[-1.1em]}}\\[-0.25em]
  \subfloat[ Case II , magnetic, $\beta=\sfrac{1}{32}$]{\parbox{0.5\textwidth}{\hspace{-0.75em}\input{beta-132.tex}\\[-1.1em]}}\\[-0.25em]
  \subfloat[ Case II , magnetic, $\beta=\sfrac{1}{16}$]{\parbox{0.5\textwidth}{\hspace{-0.75em}\input{beta-116.tex}\\[-1.1em]}}\\[-0.25em]
  \subfloat[ Case II , magnetic, $\beta=\sfrac{1}{8}$]{\parbox{0.5\textwidth}{\hspace{-0.75em}\input{beta-18.tex}\\[-1.1em]}}\\[0.2em]
  \caption{
     {Case II:} \  
    Computed eigen-frequencies $\breve\omega_n$ from the eigenvalue problem
    of Eqs.~\eqref{eq:eigenvalue_problem}
and~\eqref{eq:eigenvalue_problem_boundary_conditions-alt}. The magnetic-field related values are $\beta=0$,
    $\sfrac{1}{64}$, $\sfrac{1}{32}$, $\sfrac{1}{16}$, $\sfrac{1}{8}$; and the 
    kinetic-parameter values are $\delta=\pm\sfrac{1}{8}$,
    $\pm\sfrac{1}{16}$, $\pm\sfrac{1}{32}$, $\pm\sfrac{1}{64}$,
    $\pm\sfrac{1}{128}$, $\pm\sfrac{1}{256}$, $\pm\sfrac{1}{512}$. The
    green (red) regions contain eigen-frequencies of the stability (instability) spectrum, for $\delta>0$ ($\delta<0$). 
  }
  \label{fig:spectrum-alt}
\end{figure}
Figure~\ref{fig:spectrum} shows sequences of numerically computed,
discrete eigen-frequencies for the eigenvalue problem of 
  Case I. This problem is expressed by Eq.~\eqref{eq:eigenvalue_problem} and the boundary
conditions of Eq.~\eqref{eq:eigenvalue_problem_boundary_conditions}.
Correspondingly, Fig.~\ref{fig:spectrum-alt} shows results of our computation
for Case II. The underlying problem is described by 
Eq.~\eqref{eq:eigenvalue_problem} and the boundary
conditions of Eq.~\eqref{eq:eigenvalue_problem_boundary_conditions-alt}.
For both cases,  
the eigen-frequencies are computed for varying parameters $\beta$
and $\delta$. In Figs.~\ref{fig:spectrum} and~\ref{fig:spectrum-alt}, each plot [(a)-(e)] contains eigen-frequencies for a fixed
magnetic field with the following values: (a) $\beta=0$, (b) $\beta=\sfrac{1}{64}$, (c)
$\beta=\sfrac{1}{32}$, (d) $\beta=\sfrac{1}{16}$, and (e)
$\beta=\sfrac{1}{8}$. We relate this choice of parameter values to dimensional quantities in Sec.~\ref{subsec:discussion}.  

We compute same-$|\delta|$ pairs of eigen-frequencies of the stability spectrum (for $\delta>0$) and
the instability spectrum ($\delta<0$) with decreasing values
$|\delta|=\sfrac{1}{8}$, $\sfrac{1}{16}$, $\sfrac{1}{32}$, $\sfrac{1}{64}$,
$\sfrac{1}{128}$, $\sfrac{1}{256}$, $\sfrac{1}{512}$ for the deviation from the transition
point.  The eigen-frequencies of the stability spectrum   (for $\delta>0$) are located in
the lower half-plane since they have negative imaginary part (shaded green
regions in
 Figs.~\ref{fig:spectrum} and~\ref{fig:spectrum-alt}) ;
whereas   the eigen-frequencies of the instability spectrum   ($\delta<0$) lie in the upper
half-plane (shaded red regions in
 Figs.~\ref{fig:spectrum} and~\ref{fig:spectrum-alt}) .

 For $\beta>0$, Cases~I and~II of the lateral-flux boundary conditions yield distinct spectra. In both cases, we observe the
emergence of a \emph{gap}, defined in Sec.~\ref{subsec:gap},   across the transition point   in
the imaginary parts of the  eigen-frequencies of the stability and instability spectra.   This gap expresses an {\em
exclusion zone} (strip) that is parallel to the real axis and
separates the   stability spectrum of eigen-frequencies   ($\delta>0$) from the   instability spectrum  ($\delta<0$).  We discuss aspects of this gap in
Section~\ref{subsec:gap}. 

\begin{figure}[!]
  \centering
  \vspace{0.75em}
  \fbox{\begin{tikzpicture}[gnuplot]
\tikzset{every node/.append style={font={}}}
\gpsetlinetype{gp lt axes}
\gpsetdashtype{gp dt axes}
\gpsetlinewidth{1.00}
\gpsetpointsize{4.00}
\gpcolor{color=gp lt color border}
\node[gp node right] at (2.600,0.334) {$|\delta|=\tfrac{1}{8}$};
\gpcolor{rgb color={0.898,0.118,0.063}}
\gp3point{gp mark 7}{}{(2.640,0.334)}
\gpcolor{color=gp lt color border}
\node[gp node right] at (3.660,0.334) {$\tfrac{1}{16}$};
\gpcolor{rgb color={0.000,0.447,0.698}}
\gp3point{gp mark 6}{}{(3.700,0.334)}
\gpcolor{color=gp lt color border}
\node[gp node right] at (4.720,0.334) {$\tfrac{1}{32}$};
\gpcolor{rgb color={0.941,0.894,0.259}}
\gp3point{gp mark 5}{}{(4.760,0.334)}
\gpcolor{color=gp lt color border}
\node[gp node right] at (5.780,0.334) {$\tfrac{1}{64}$};
\gpcolor{rgb color={0.902,0.624,0.000}}
\gp3point{gp mark 4}{}{(5.820,0.334)}
\gpcolor{color=gp lt color border}
\node[gp node right] at (6.840,0.334) {$\tfrac{1}{128}$};
\gpcolor{rgb color={0.337,0.706,0.914}}
\gp3point{gp mark 3}{}{(6.880,0.334)}
\gpcolor{color=gp lt color border}
\node[gp node right] at (7.900,0.334) {$\tfrac{1}{256}$};
\gpcolor{rgb color={0.000,0.620,0.451}}
\gp3point{gp mark 2}{}{(7.940,0.334)}
\gpcolor{color=gp lt color border}
\node[gp node right] at (8.960,0.334) {$\tfrac{1}{512}$};
\gpcolor{rgb color={0.580,0.000,0.827}}
\gp3point{gp mark 1}{}{(9.000,0.334)}
\gpdefrectangularnode{gp plot 1}{\pgfpoint{0.460cm}{0.676cm}}{\pgfpoint{13.417cm}{3.501cm}}
\end{tikzpicture}}\\[+0.25em]
  {\input{nbc-extraneous.tex}}
  \vspace{-1.5em}
  \caption{
     Case I: Deviation of the lowest positive eigen-frequency, $\breve \omega_0^e(\beta,
    \delta)$, of the extraneous spectrum from its non-magnetic value $\breve \omega_0^{e,0}=\breve \omega_0^e(0,
    \delta)=\pi$ as a function of
    $\mathrm{sgn}(\delta)\beta$ ($\beta>0$). The numerical values of the kinetic parameter are
    $\delta=\pm\sfrac{1}{8}$, $\pm\sfrac{1}{16}$, $\pm\sfrac{1}{32}$,
    $\pm\sfrac{1}{64}$, $\pm\sfrac{1}{128}$, $\pm\sfrac{1}{256}$,
    $\pm\sfrac{1}{512}$. As $|\delta|$ decreases, $\breve \omega_0^e(\beta,
    \delta)-\pi$ asymptotically approaches a straight line (dashed line).  }
  \label{fig:extraneous}
\end{figure}

For Case~I,  we notice that a part of the numerically computed spectrum is   real valued and   distinct from the DS prediction if $\beta=0$.   This part, which remains real valued even when $\beta$ is nonzero, is referred to as the {\em extraneous spectrum}.  We provide a justification for this characterization in Sec.~\ref{subsec:discussion}. These real eigen-frequencies are denoted by $\breve\omega_n^e(\beta,\delta)$, and are continuous with $\beta$; thus, there is no growth or decay associated with these frequencies. In Appendix~\ref{app:limit-zero-B}, we derive this extraneous spectrum in the limit $\beta\to 0$ of our magnetically driven model.     For $\beta=0$, the extraneous eigen-frequencies turn out to be $\delta$-independent, $\breve\omega_n^e(0,\delta)=\breve\omega_n^{e,0}=(2n+1)\pi$ with $n\in \mathbb{Z}$ (see Appendix~\ref{app:limit-zero-B}). In Fig.~\ref{fig:extraneous}, we plot the deviation of the lowest positive eigenvalue, $\breve\omega_0^e(\beta,\delta)$, of the extraneous spectrum from its non-magnetic counterpart, $\breve\omega_0^{e,0}=\pi$, as a function of $\mathrm{sgn}(\delta)\beta$ for different values of $\delta$ ($\beta>0$). Note that there is no eigen-frequency of the extraneous spectrum shown in Fig.~\ref{fig:spectrum},
as all these frequencies lie outside of our plotting domain.
   In Case~II, the non-magnetic limit of the spectrum is  identical to that of the DS model~\cite{DyakonovShur1993}.   In contrast, in Case~I the resulting (non-magnetic) spectrum is the union of two sets.    One part is the (complex-valued) DS spectrum. The other, real valued, extraneous part is due to  a geometric resonance inherent to the lateral-flux boundary condition.   This spectrum is accompanied by  sinusoidal lateral fluxes    of  wavelengths $L/|n+\tfrac{1}{2}|$ (Appendix~\ref{app:limit-zero-B}).

\subsection{Anatomy of spectral gap}
\label{subsec:gap}
\begin{figure}[!]
  \centering
  \subfloat[ Case I ]{\input{nbc-gap.tex}}
  \\
  \subfloat[ Case II ]{\input{gap.tex}}
  \caption{
    Spectral distance $D_{\breve\omega}(\beta,\delta)$, given by
    Eq.~\eqref{eq:spectral-dist-def-computational}, as a function of
    kinetic parameter $|\delta|$, for several values of $\beta$
    ($0\le\beta\ll 1$),  for Case I [(a)] and Case II
    [(b)].   Dashed red curve: The respective plot of
    $D_{\tilde\omega}^{\text{DS}}(\delta)$, for the reference case of the DS model, by use of Eq.~\eqref{eq:spectral-dist-def-DS}.
  }
  \label{fig:gap}
\end{figure}
\begin{figure}[!]
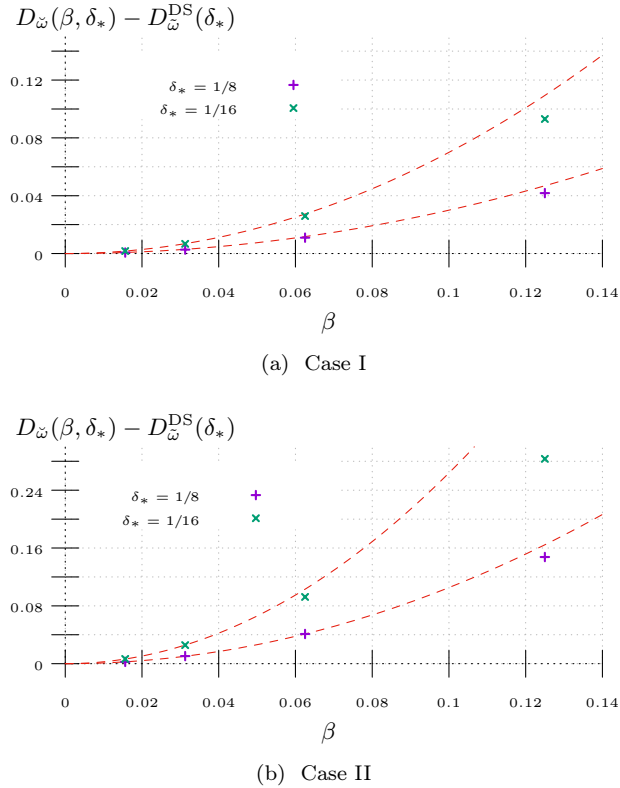

  \centering
  \subfloat[ Case I ]{\input{nbc-square.tex}}
  \\
  \subfloat[ Case II ]{\input{square.tex}}
  \caption{
    Numerical evidence for scaling law~\eqref{eq:beta-scaling-eigenf}: Computed $D_{\breve\omega}(\beta,\delta_*) -
    D_{\tilde\omega}^{\text{DS}}(\delta_*)$ versus $\beta$ with $\beta
    \breve l\ll \sqrt{2|\delta|}$, for fixed values
    $\delta_*=\sfrac{1}{8}$ and $\delta_*=\sfrac{1}{16}$,  
    for Case I [(a)] and Case II [(b)]. Dashed red curve: Best fit to a parabola of form $C \beta^2$, in view of Eq.~\eqref{eq:beta-scaling-eigenf}.  
  }
  \label{fig:square}
\end{figure}

Next, we quantitatively define  the notion of the spectral gap mentioned
above. Consider a given, fixed kinetic parameter $|\delta|$ and magnetic
field $\beta>0$. We define the \emph{spectral distance},
$D_{\breve\omega}(\beta,\delta)$, as follows: Subtract the (strictly negative)
maximum of the imaginary parts of all computed eigen-frequencies
of the  stability spectrum   
from the (strictly positive) minimum of the imaginary parts of all 
eigen-frequencies of the  instability spectrum. This operation implies the following formula:  
\begin{multline}
  \label{eq:spectral-dist-def-computational}
  D_{\breve\omega}(\beta,\delta) =
  \min_n\big\{\imp\,\breve\omega_n(-|\delta|) >0 \big\}
  \\
  -
  \max_n\big\{\imp\,\breve\omega_n(|\delta|) <0 \big\}.
\end{multline}
This definition of the spectral distance is an extension of the spectral
distance $D_{\tilde\omega}^{\text{DS}}(\delta)$ introduced in
Eq.~\eqref{eq:spectral-dist-def-DS}, for the DS model.  Recall the related scaled eigen-frequencies $\tilde\omega_n$ of Eq.~\eqref{eq:scaled-omega-DS}.
The definition of Eq.~\eqref{eq:spectral-dist-def-computational} excludes the extraneous spectrum, which is present in
Case I.  

In Fig.~\ref{fig:gap}, we show the effects of the kinetic deviation parameter $\delta$ and magnetic field $\beta$ on the
computed spectral distance, $D_{\breve\omega}(\beta,\delta)$, for the values $\beta=0,\,
\sfrac{1}{64},\,\sfrac{1}{32},\,\sfrac{1}{16},\,\sfrac{1}{8}$. For the sake
of comparisons, in Fig.~\ref{fig:gap} we also plot the function
$D_{\tilde\omega}^{\text{DS}}(\delta)$ which signifies the non-magnetic
limit, when $\beta=0$, corresponding to the DS model (dashed red curve). We verify that the numerically
computed spectral distance with $\beta=0$ agrees with
formula~\eqref{eq:spectral-dist-def-DS} for
$D_{\tilde\omega}^{\text{DS}}(\delta)$; this function is monotone in
$\delta$ with $D_{\tilde\omega}^{\text{DS}}(\delta)\to0$ as $\delta\to0$.

For $\beta>0$, we define the \emph{spectral gap},
$G_{\breve \omega}(\beta)$, as follows:
\begin{align}
  G_{\breve\omega}(\beta)\;=\;
  \min_{\delta>0}D_{\breve\omega}(\beta,\delta)\;>\;0,
  \label{eq:gap_definition}
\end{align}
which is strictly positive in the presence of a (nonzero) magnetic field.
An inspection of Fig.~\ref{fig:gap} suggests that
 for Case I we have
\begin{align*}
  \begin{aligned}
    G_{\breve\omega}(\sfrac1{8}) &\approx 0.41, &\quad
    G_{\breve\omega}(\sfrac1{16})&\approx 0.23,
    \\
    G_{\breve\omega}(\sfrac1{32})&\approx 0.13, &\quad
    G_{\breve\omega}(\sfrac1{64})&\approx 0.07.
  \end{aligned}
\end{align*}
For Case II, the corresponding values are  
\begin{align*}
  \begin{aligned}
    G_{\breve\omega}(\sfrac1{8}) &\approx 0.70, &\quad
    G_{\breve\omega}(\sfrac1{16})&\approx 0.41,
    \\
    G_{\breve\omega}(\sfrac1{32})&\approx 0.24, &\quad
    G_{\breve\omega}(\sfrac1{64})&\approx 0.13.
  \end{aligned}
\end{align*}
 These observations hint \ at a (sub-)linear dependence of the gap on
$\beta$. Lastly, we   provide numerical evidence for   the heuristically derived
$\beta^2$-scaling law regarding the deviation of the eigen-frequencies
from their non-magnetic limit   of the DS model   under an extremely weak magnetic field
(Sec.~\ref{sssec:asymptotics-ss}). To this end, we compute the difference
$D_{\breve\omega}(\beta,\delta) - D_{\tilde\omega}^{\text{DS}}(\delta)$ for
fixed $\delta=\delta_*$ and as a function of the extremely weak magnetic
field, when $\beta\breve l\ll \sqrt{2|\delta|}$. The numerical results are
displayed in Fig.~\ref{fig:square}. Evidently, the scaling prediction of
Eq.~\eqref{eq:beta-scaling-eigenf} ceases to follow the numerically
computed values if $\beta^2$ becomes sufficiently large.

\subsection{Discussion of results}
\label{subsec:discussion}

Although our computations are restricted to relatively weak magnetic fields
in the linear regime of the shallow-water equation model, the results may
have more general implications. We believe that we predict trends that
persist even in cases with large magnetic perturbations, away from the
critical transition point, $b=\sfrac1{2}$ (or, $\delta=0$). 

 There are two main predictions indicated by our numerics.
Notably, in the instability regime the growth exponents increase with
$\beta$; and a spectral gap is present near the transition point. We
discuss these observations below (Sec.~\ref{sssec:Gap_exp}). In addition,
we discuss the meaning of the extraneous spectrum for Case~I
(Sec.~\ref{sssec:extran-spec}).

\subsubsection{Growth exponents and spectral gap}
\label{sssec:Gap_exp}
It is of interest to relate the values of the dimensionless parameters used
in our numerical computations to dimensional physical quantities, in the
context of the shallow-water equations. Let $s_0/L$ and $\bar m_e$ (ratio of
effective mass $m_e$ to free electron mass $m_{e0}$) be free parameters,
without specifying the 2D material. For $\breve l=1$, by the definitions of
$B_0$ and rescaled frequency $\breve \omega$ we find 
\begin{equation*}
  \omega_n=\breve \omega_n \sqrt{1+2\delta}\,\frac{s_0}{L},\quad
  B_0=\frac{m_{e0}}{e}{\bar m}_e\,\frac{s_0}{L},
\end{equation*}
with $B= \beta B_0$. Note that $B_0$, and thus $B$ for given $\beta$, is
proportional to $\bar m_e$. For the values of $\beta$ and $\delta$ in our
numerics, the real parts of the computed frequencies $\breve \omega_n$ may
lie in the appealing range of THz if $s_0/L$ takes suitable values. For
example, consider $s_0/L$ between $0.5$ THz and $10$ THz. Consequently, we
would have $2.8 \bar m_e\,\textrm{T} \lesssim B_0 \lesssim 57 \bar
m_e\,\textrm{T}$, which amounts to
\begin{equation}\label{eq:conversion}
  0.5\breve\omega_n\,\text{THz}
  \lesssim\frac{\omega_n}{\sqrt{1+2\delta}}\lesssim 10
  \breve\omega_n\,\text{THz}
\end{equation}
where $-1/2<\delta<0$ in the subsonic regime.
Relation~\eqref{eq:conversion} does not involve $\bar m_e$; and may be used
near the transition point, for small $|\delta|$, if this kinetic regime is
accessible. For $\beta=\sfrac1{8}$, the largest value of $\beta$ in our
numerics, we obtain $0.35\bar m_e\,\text{T}\lesssim B\lesssim 7\bar
m_e\,\text{T}$. For fixed channel width, e.g., $L=1\,\mu\text{m}$, a range
for $s_0$ ensues, e.g., $0.5\times10^6\,\textrm{m/s}\lesssim s_0 \lesssim
10^7\,\textrm{m/s}$.

Our discussion here relies on the assumption that the indicated values of
$\delta$ are physically accessible within the model of the Euler equations.
We have been unable to propose with certainty a specific 2D material that
can serve this purpose when $|\delta|$ is small. Hence, the feasibility of
observing the predicted  spectral gap is left open.

On the other hand, an inspection of Figs.~\ref{fig:spectrum},
\ref{fig:spectrum-alt}, and~\ref{fig:square} suggests that the growth
exponents, i.e., the imaginary parts of the eigen-frequencies in the
instability spectrum, increase with $\beta$ away from the critical point. This prediction, if extended deeply into the subsonic regime (for $\delta$ close to $-1/2$), may plausibly be testable in 2D materials already proposed for the DS instability; see, e.g.,~\cite{MendlLucas2018}.  

Therefore, for fixed channel width and Mach number of the imposed boundary
conditions, the manifestation of the instability is plausibly enhanced with
the increasing magnitude of the static, out-of-plane magnetic field. The
enhancement is evident from a small magnetic perturbation of the
governing equations of motion in our approach. This trend should persist
even for larger magnetic fields within our model. It is tempting to expect
that, if a DS-type instability is experimentally feasible in a 2D
electronic fluid, it can be  amplified by an applied static magnetic field.
This hypothesis is open to further model refinements, and experimental
verification.

The numerically observed gap
between the stability ($\delta>0$) and instability ($\delta<0$) spectra, viewed as a {\em feature of a particular model} for the electronic fluid, may challenge the  realization of a DS-type instability.  
This property suggests that there is no continuous passage from stability to
instability through the variation of the Mach number within the shallow-water equations.  Further scrutiny of this regime might require modification of the model. This task is not pursued in this paper.

\subsubsection{Comparison of Cases I and II, and extraneous spectrum}
\label{sssec:extran-spec}
Let us first review the stability formulation. Each of Cases~I and~II has a unique steady state. In Case~I, the steady state yields lateral fluxes $\pm eB L\rho_0/(2m_e)=\pm \beta \rho_0s_0/2$ (for $\breve l=1$) at the left and right boundaries. As $\beta\to 0$, each steady state reduces to the DS case of constant density and normal flux, and zero tangential flux everywhere. We perturb the system around the steady states by linearizing the Euler equations with respective homogeneous boundary conditions; and compute the stability/instability spectra via eigenvalue problems. For $\beta\neq 0$ the spectra, and their sets of eigenmodes, are distinct for Cases~I and~II. In Case~I, for all $\beta$ the solution of the lateral-flux perturbation can {\em in principle} attain arbitrary nonzero yet mutually opposite values at the boundaries, which depend on the eigenmode. As $\beta\to 0$, the spectrum of Case~I reduces to the union of the DS spectrum and a set of real frequencies, referred to as the extraneous spectrum; while Case~II yields only the DS spectrum. In this non-magnetic limit, for Case~I the extraneous spectrum yields zero density and normal-flux  perturbations  whereas the remaining (DS) spectrum entails zero lateral-flux perturbation everywhere in the channel, as shown in Appendix~\ref{app:limit-zero-B}. The extraneous spectrum remains real valued if $\beta\neq 0$. 

The superposition of eigenmodes for each case can express the unique solution of the {\em time dependent} system, Eq.~\eqref{eq:scaled-model_with_B}, by linearization of the motion laws and imposition of suitable initial conditions under the same boundary conditions. Hence, the physically relevant eigenmodes and their contributions should be determined from the {\em initial preparation} of the system. For example, in Case~I it is natural to expect and impose that, as $\beta\to 0$, the lateral-flux perturbation vanishes initially. The details of the time-dependent problem lie beyond our scope.

Let us now discuss the meaning of the extraneous spectrum (Case~I). We reiterate two main features. First, this spectrum is real valued; thus, there is no growth (or decay) associated with it. Since the electron flow cannot be unstable at these frequencies, this response can hardly be viewed as a mechanism of generating THz electromagnetic radiation in the DS sense~\cite{DyakonovShur1993}. Second, in the limit $\beta\to0$ the eigenmodes of the extraneous spectrum yield vanishing perturbations of density and normal flux, and a sinusoidal lateral-flux perturbation everywhere; see Appendix~\ref{app:limit-zero-B}. For an actual physical setting with $\beta=0$, we can assume that the density and normal flux are {\em initially} perturbed around the steady state {\em nontrivially}, with nonzero values, while the lateral flux is zero. The imposition of such initial conditions would imply the elimination of the eigenmodes with real frequencies in the time-dependent linear response. By continuously perturbing this non-magnetic setting with $\beta$, we can connect the ensuing time-dependent solution only to the complex, non-real valued spectrum of Case~I for $\beta\neq 0$. Thus, the extraneous spectrum becomes irrelevant.


\section{Conclusion}
\label{sec:conclusion}

In this paper, we studied the effect of a static, out-of-plane magnetic
field on the stability and instability spectra of a 2D inviscid electronic
fluid  in an infinitely long channel.   Our model
forms an extension of the Dyakonov-Shur model of plasma wave
generation in 1D.   We considered two cases of the boundary condition
for the lateral flux. One of these cases (Case~I)--with opposite lateral fluxes at the boundaries--is deemed as physically plausible. Another case (Case~II)--with a zero lateral flux at the right boundary--is studied for comparison purposes.   The main results of our work can be summarized as
follows.  First, we derived closed-form analytical expressions for the respective
spatially dependent steady state; and formulated  eigenvalue problems by
linear perturbations.  Second, we demonstrated by numerics
that the magnetic field enhances the growth exponents of the fluid
instability. Third, we provided numerical evidence for the emergence of a
magnetically controlled spectral gap. 

 Our work also indicates a subtlety inherent to the DS model. The DS
spectrum may not always arise uniquely in the limit of a zero magnetic
field in 2D. As an example, we showed that the imposition of mutually opposite
boundary values of the lateral flux yields two distinct spectra, only one of which coincides with the DS prediction in each 
kinetic regime in the non-magnetic limit.   The other spectrum contains only real frequencies. This latter spectrum is deemed as irrelevant for our purpose, since it cannot induce an unstable electron flow.

Some open questions should be noted.   In an
actual 2D experimental setup, the channel has a finite length, say, $L_y$. This geometry breaks translation invariance in
$y$, and requires suitable conditions  at the top
and bottom boundaries. For a broad family of such conditions,
our results for the steady state would approximately hold away from the top and bottom
boundaries if the channel is long enough, $L_y\gg L$. We
expect that the lowest part of our computed spectrum would provide a good
zeroth-order approximation if the length scale over which the fields of
those modes vary, which is of the order of $|v_{x0}+s_0|/|\omega_n|$, is
small compared to $L_y$.   The combined
effect of nonlinearities and magnetic field on the growth exponents has
not been addressed.     Finally, our prediction of a
spectral gap could be the subject of laboratory testing, as well as the motivation for studying the validity or modification of the hydrodynamic model near the transition point.


\begin{acknowledgements}
  The authors are indebted to Prof.~Andrew Lucas for bringing
  Ref.~\cite{DyakonovShur1993} and related published works to their
  attention, as well as for discussing physical aspects of the DS model
  with them. The authors also wish to thank 
  Prof.~Mitchell Luskin and
  Prof.~Huy Q. Nguyen for useful discussions on the DS model;
    and an anonymous reviewer for suggesting the condition of
  mutually opposite boundary values of the tangential flux.  
  The work of the first author (M.M.) was supported in part by the National
  Science Foundation grants DMS-1912847, DMS-2045636, and by the Air Force
  Office of Scientific Research, USAF, under grant/contract number
  FA9550-23-1-0007. The work of the second author (D.C.-R.) was supported
  by the NSF-REU via Grant No. DMS-2149913 at the University of Maryland
  during the summer of 2022. The third author (D.M.) is grateful to the
  School of Mathematics of the University of Minnesota for hosting him as
  an Ordway Distinguished Visitor in the spring of 2022, when part of this
  work was carried out.
\end{acknowledgements}


\appendix
\section{Steady-state solutions for $\beta\neq 0$}
\label{app:steady_state}

In this appendix,  we derive the
exact steady-state solutions of
Sec.~\ref{subsec:steady_state}.   
We also extract small-$\beta \breve l$ approximate formulas for the
electronic number density.

\subsection{Derivation of steady states for Cases~I and~II}

 Let us first address the case with mutually opposite lateral
fluxes at the two boundaries, Case~I.   By using
Eq.~\eqref{eq:scaled-model_with_B} with $(\partial/\partial t)(\breve\rho,
\breve v_x, \breve v_y)=0$ and the notation $(\breve\rho, \breve v_x,
\breve v_y)=(\breve\rho^s, \breve v_x^s, \breve v_y^s)$ , we obtain
\begin{align}
  \label{eq:app:scaled-model_with_B-ss}
  \begin{rcases*}
    \begin{aligned}
      \breve v_x^s\frac{\partial \breve v_x^s}{\partial \breve
      x}+(2b)^{-1}\,\frac{\partial\breve \rho^s}{\partial \breve x}
      \;&=\; \beta (2b)^{-1}\,\breve v_y^s,
      \\
      \breve v_x^s\frac{\partial\breve v_y^s}{\partial \breve x}
      \;&=\; -\beta \breve v_x^s,
      \\
     \frac{\partial}{\partial \breve x}(\breve \rho^s\,\breve v_x^s)\;&=\,0.
    \end{aligned}
  \end{rcases*}
\end{align}
This is a system of ordinary differential equations (ODEs) subject to the
conditions   of Eq.~\eqref{eq:scaled-BCs-opposite}.  
The last two ODEs are integrated directly, yielding (if $\breve v_x^s\neq
0$)
\begin{align}
  \label{eq:app:vy-rhovx-ss}
  \begin{rcases*}
    \begin{aligned}
      \breve v_y^s&= \beta \big(\tfrac{\breve l}{2}-\breve x\big),
      \\
      \breve\rho^s\, \breve v_x^s&= 1,\qquad 0\le x\le \breve l.
    \end{aligned}
  \end{rcases*}
\end{align}
Hence, by Eq.~\eqref{eq:app:scaled-model_with_B-ss} the Euler equation for
$\breve v_x^s$ reduces to
\begin{align*}
  b\frac{\partial}{\partial \breve x}(\breve v_x^s)^2+\frac{\partial\breve
  \rho^s}{\partial \breve x}=\beta^2  \big(\tfrac{\breve l}{2}-\breve x\big),  
  \quad 0\le \breve x\le \breve l.
\end{align*}
This equation is integrated by application of the condition
$\breve\rho^s=1$ at $\breve x=0$, and furnishes
\begin{equation*}
  \breve \rho^s-1 +b [(\breve{v}_x^s)^2-1]=\tfrac{\beta^2}{2}
   \breve x(\breve l -\breve x),  
  \quad 0\le \breve x\le \breve l.
\end{equation*}
Recalling that $\breve\rho^s \,\breve v_x^s=1$, we find the cubic equation
\begin{equation}
  \label{eq:app:cubic-zeta}
  \zeta^3-\zeta^2+\epsilon^2=0,
\end{equation}
where $\zeta=[1+b+(\beta^2/2)  \breve x(\breve l-\breve x) ]^{-1}
\breve\rho^s$ and
\begin{equation}
  \label{eq:app:epsilon2-form}
  \epsilon^2= \frac{b}{\big[1 + b + \tfrac{\beta^2}{2}  \breve
  x(\breve{l} - \breve{x} )\big]^3}.
\end{equation}
Note that  $b(1+b+\beta^2\breve l^2/8)^{-3}\le \epsilon(\breve
x)^2\le b(1+b)^{-3}$   for $0\le  \breve x\le \breve l$, which
precisely describes the minimum and the maximum value of $\epsilon(\breve
x)^2$ in the channel. The maximum value in $\breve x$ is attained if we let
$\breve x=0$  or $\breve x=\breve l$,   and (as a
function of $b$) is maximized, becoming equal to $4/27$, for $b=1/2$. By
writing Eq.~\eqref{eq:app:cubic-zeta} as $\epsilon^2=\zeta^2(1-\zeta)$, we
notice that any physically admissible solution, $\zeta=\zeta^s$, must
satisfy $0< \zeta^s\le 1$ for $b> 0$. The value $\zeta^s=(1+b)^{-1}$ is an
admissible root for $\epsilon^2=b(1+b)^{-3}$,   which corresponds
to the boundary condition for $\breve\rho^s$ at $\breve x=0$ and gives the
same value at $\breve x=\breve l$.  

Next, we solve Eq.~\eqref{eq:app:cubic-zeta} by the known procedure based
on Vieta's substitution~\cite{BirkhoffMacLane,AbramowitzStegun-Handbook}.
This equation is of the form $\zeta^3+a_2 \zeta^2+a_1 \zeta+a_0=0$ with
$a_2=-1$, $a_1=0$ and $a_0=\epsilon^2$. Accordingly,
following~\cite{AbramowitzStegun-Handbook} we define
\begin{align*}
  q&=\tfrac{1}{3}a_1-\tfrac{1}{9}a_2^2=-\frac{1}{9},
  \\
  p&=\tfrac{1}{6}(a_1 a_2-3a_0)-\tfrac{1}{27}a_2^3
  =-\frac{\epsilon^2}{2}+\frac{1}{27}.
\end{align*}
The sign of the quantity $q^3+p^2$ determines how many real roots the cubic
equation has. We compute
\begin{equation*}
  q^3+p^2=\frac{\epsilon^2}{4}\left(\epsilon^2-\frac{4}{27}\right)\le 0.
\end{equation*}
Hence, all roots of Eq.~\eqref{eq:app:cubic-zeta} are
real~\cite{BirkhoffMacLane,AbramowitzStegun-Handbook}. Now define
\begin{align*}
  \varsigma_\pm=[p\pm (q^3+p^2)^{1/2}]^{1/3},
\end{align*}
with the typical convention that $\imp\,(q^3+p^2)^{1/2}\ge 0$. The root of
interest is written as
\begin{align}
  \zeta^s&=\varsigma_+ +\varsigma_--\frac{a_2}{3}\notag
  \\
  &=\frac{1}{3}+\sum_{s=\pm 1} \biggl(\frac{1}{27}-\frac{\epsilon^2}{2} +
  \im\,s \frac{\epsilon}{2}\sqrt{\frac{4}{27}-\epsilon^2}\biggr)^{1/3}.
\end{align}
This root is the only physically admissible one. For example, it reduces to
the solution of the non-magnetic case, as $B\to 0$~\cite{DyakonovShur1993}.
We omit the details of  this limit here.

To simplify the above expression for $\zeta^s$, we use the following
identity, for all real $\epsilon$ with $|\epsilon|\le \frac{2}{3\sqrt{3}}$:
\begin{equation*}
  \left|\frac{1}{27}-\frac{\epsilon^2}{2}\pm\im
  \frac{\epsilon}{2}\sqrt{\frac{4}{27}-\epsilon^2}\right|=\frac{1}{27},
\end{equation*}
by which we set
\begin{equation*}
  \frac{1}{27}-\frac{\epsilon^2}{2}\pm\im
  \frac{\epsilon}{2}\sqrt{\frac{4}{27}-\epsilon^2}=\frac{1}{27}\,e^{\pm
  \im\theta},\quad 0\le \theta\le 2\pi.
\end{equation*}
Thus, we obtain the anticipated formula
\begin{equation}\label{eq:app:zeta-theta}
  \zeta^s=\frac{2}{3}\cos\biggl(\frac{\theta}{3}\biggr)+\frac{1}{3},
\end{equation}
where
\begin{equation}\label{eq:app:theta-orig}
  \cos\theta=1-\frac{27}{2}\epsilon^2.
\end{equation}
This equation must be inverted for $\theta$ in view of
Eq.~\eqref{eq:app:zeta-theta}. Thus, $\theta$ is given in terms of some
\emph{generalized} inverse cosine,  a multivalued function,
  with argument $1-(27/2)\epsilon^2$, which must be evaluated
by appropriate analytic continuation in $\epsilon^2$ as $\breve x$ and $b$
vary. Once $\zeta^s$ is determined in this way, we have
\begin{equation}\label{eq:app:rho-ss-total}
  \breve \rho^s=[1+b+(\beta^2/2) \breve x(\breve l-\breve x) ]
  \left[\frac{2}{3}\cos\biggl(\frac{\theta}{3}\biggr)+\frac{1}{3}\right]
\end{equation}
and $\breve v_x^s=1/\breve\rho^s$ for all $\breve x$ ($0\le \breve x\le
\breve l$).

We should outline the implications from the multivaluedness that is
inherent to the generalized inverse cosine function, as $\epsilon^2$ varies
through the change of $b$ and $\breve x$ ($0\le \breve x\le \breve l$). For
definiteness, we invoke the principal branch of $\cos^{-1}(\cdot)$, defined
by
\begin{equation*}
  0\le \cos^{-1}z\le \pi,\quad \mbox{if}\ -1\le  z\le 1.
\end{equation*}
From now on, the symbol $\cos^{-1} z$ denotes a real single-valued function
under the above definition.

First, consider the solution with
\begin{equation}\label{eq:app:theta-branch0}
  \theta(\breve x;b)=\cos^{-1}\biggl(1-\frac{27}{2}
  \epsilon(\breve x;b)^2\biggr)~,
\end{equation}
which entails $2/3 <\zeta^s\le 1$ by virtue of
Eq.~\eqref{eq:app:zeta-theta}. The boundary condition at $\breve x=0$ reads
as $\zeta^s=(1+b)^{-1}$, and is manifest in this particular branch only if
\begin{equation*}
  0\le b < 1/2
\end{equation*}
which signifies the subsonic regime. For this range of $b$, if $\breve x$
increases continuously  from $0$ to $\breve l$ the quantity
$1-(27/2)\epsilon(\breve x)^2$ increases continuously taking values in an
interval contained entirely in $[-1, 1]$. We thus verify (self
consistently) that in this kinetic regime $\theta(\breve x;b)$ remains in
the same branch for the whole channel up to the boundaries, $0\le \breve x
\le\breve l$.

On the other hand, consider the solution with
\begin{equation}\label{eq:app:theta-branch1}
  \theta(\breve x;b)=2\pi-\cos^{-1}\biggl(1-\frac{27}{2}
  \epsilon(\breve x;b)^2\biggr)~,
\end{equation}
which corresponds to a different branch of the underlying generalized
inverse cosine  and entails $0\le  \zeta^s < 2/3$. Accordingly, the
enforcement of the boundary condition at $\breve x=0$ is equivalent to the
restriction
\begin{equation*}
  b> 1/2,
\end{equation*}
which characterizes the supersonic regime. As $\breve x$ continuously
increases from $0$ to $\breve l$ inside the channel, we verify that the
function $\theta(\breve x;b)$  lies in this particular branch.

Hence, for fixed $\breve x$  ($\breve x\neq 0,\,\breve l$)   and nonzero $\beta$,   
 we can show that   the continuous variation of $b$ from values $b<1/2$ to $b>1/2$, through
the transition point, reveals a jump discontinuity of the electronic number
density (as a function of $b$) at $b=1/2$.  The jump disappears in the
limit of zero magnetic field ($\beta\to 0$), when the density approaches a
constant everywhere in the channel~\cite{DyakonovShur1993}.

 Let us now discuss the case with a vanishing lateral flux at the right boundary, Case~II. We need to solve
Eq.~\eqref{eq:app:scaled-model_with_B-ss} subject to the conditions of
Eq.~\eqref{eq:scaled-BCs}. We obtain
\begin{align}
  \label{eq:app:vy-rhovx-ss-alt}
  \begin{rcases*}
    \begin{aligned}
      \breve v_y^s&=\beta \big(\breve l-\breve x\big),
      \\
      \breve\rho^s\, \breve v_x^s&= 1,\qquad 0\le x\le \breve l.
    \end{aligned}
  \end{rcases*}
\end{align}
Thus, the Euler equation for $\breve v_x^s$ becomes
\begin{align*}
  b\frac{\partial}{\partial \breve x}(\breve v_x^s)^2+\frac{\partial\breve
  \rho^s}{\partial \breve x}=\beta^2 \big(\breve l-\breve x\big), 
  \quad 0\le \breve x\le \breve l.
\end{align*}
We find Eq.~\eqref{eq:app:cubic-zeta} with $\zeta=[1+b+(\beta^2/2) \breve x(2\breve l-\breve x)]^{-1}
\breve\rho^s$ and
\begin{equation}
  \label{eq:app:epsilon2-form-alt}
  \epsilon^2= \frac{b}{\big[1 + b + \tfrac{\beta^2}{2} \breve
  x(2\breve{l} - \breve{x})\big]^3}.
\end{equation}
 Note that $b(1+b+\beta^2\breve l^2/2)^{-3}\le \epsilon(\breve x)^2\le
b(1+b)^{-3}$, where the maximum value of $\epsilon(\breve x)^2$ is attained
at $\breve x=0$.   Without further ado, we recover
formula~\eqref{eq:app:rho-ss-total} with the replacement of the factor
$\breve x (\breve x-\breve l)$ by $\breve x(2\breve l-\breve x)$. The
function $\theta(\breve x;b)$ is defined by
Eq.~\eqref{eq:app:theta-branch0} if $0\le b< 1/2$, or
Eq.~\eqref{eq:app:theta-branch1} if $b>1/2$ in view of
Eq.~\eqref{eq:app:epsilon2-form-alt}. Thus, the branching described in
Case~I above persists here as well.  

\subsection{Small-$\beta\breve l$ perturbations}

Next, we simplify formula~\eqref{eq:app:rho-ss-total} when the magnetic
field is sufficiently weak,  focusing on  Case~I, with 
mutually opposite lateral fluxes at the boundaries.   For fixed $b\neq \tfrac{1}{2}$, the
density $\breve \rho^s$ is controlled by the parameter $\epsilon(\breve
x)^2$. By Eq.~\eqref{eq:app:epsilon2-form}, we notice that
$\epsilon^2$ varies spatially according to the term $\tfrac{\beta^2}{2} \breve x(\breve l-\breve
x)$ which is positive and less than or equal to $\mathcal
O\big((\beta\breve l)^2\big)$ if $0<\breve x< \breve l$. Thus, the relevant parameter is $\beta\breve l$. We enforce the condition $|\beta|\breve l\ll 1$.

First, let us apply regular perturbations in $\beta$. To reduce the
algebra, we write the cubic equation for $\breve\rho^s$ as
\begin{equation*}
  1-\breve \rho^s=\frac{\beta^2}{2} \breve x (\breve l-\breve
  x) \frac{(\breve \rho^s)^2}{b(1+\breve \rho^s)-(\breve \rho^s)^2}.
\end{equation*}
The right-hand side of this equation is small if $b$ is away from the
transition point. In this case, an expansion for $\breve\rho^s$ can be
obtained by iterations with the zeroth-order solution $\breve\rho^s=1$. The
first iteration gives
\begin{equation*}
  1-\breve \rho^s= -\frac{\beta^2}{2(1-2b)} \breve x (\breve l-\breve x) +
  \mathcal O\big((\beta\breve l)^4\big),
\end{equation*}
which entails Eq.~\eqref{eq:scaled-rho-ss-smallB-reg}. This expansion is
meaningful if
\begin{equation*}
  \frac{\beta^2}{2(1-2b)} \breve x (\breve l-\breve x)\ll 1.  
\end{equation*}
Thus, we must have $(\beta\breve l)^2\ll |1-2b|$.

When $(\beta\breve l)^2=\mathcal O(|1-2b|)$, the above regular expansion
for $\breve \rho^s$ breaks down. In this case, $b$ is near $1/2$. Define
the small parameters
\begin{equation}\label{eq:varepsilon-def}
  \delta=b-\frac{1}{2},\quad
  \varepsilon=\frac{\beta^2}{2} \breve x(\breve l-\breve x).
\end{equation}
We are interested in the regime with $\varepsilon\le \mathcal O(\delta)$.
Let $\breve \rho^s=1+\varrho_c$. The correction term, $\varrho_c$,
satisfies
\begin{equation*}
  -2(\delta+\varepsilon)\varrho_c+\left[\tfrac{3}{2}-
  (\delta+\varepsilon)\right]\varrho_c^2+\varrho_c^3=\varepsilon.
\end{equation*}
For the critical scaling of interest, we set $\delta+\varepsilon=\alpha
\delta$ for some $\alpha=\mathcal O(1)$. By dominant balance, we find
\begin{equation*}
  \varrho_c^2\simeq \frac{2\varepsilon}{3}\,\Rightarrow \,
  \varrho_c \simeq \pm\sqrt{\frac{2\varepsilon}{3}}.
\end{equation*}
By Eq.~\eqref{eq:app:rho-ss-total} for $\theta\simeq \pi$ and $(\beta\breve
l)^2\ll 1$, we assert that $\breve\rho^s-1>0$ for $b<1/2$ (if $\theta<\pi$)
and $\breve\rho^s-1<0$ for $b>1/2$ ($\theta>\pi$) near the transition point
($b\simeq 1/2$). Thus, we must set $\pm = \mathrm{sgn}(1-2b)$ in the
approximate formula for $\varrho_c$. This substitution yields
Eq.~\eqref{eq:scaled-rho-ss-smallB-sing}.

 Alternatively, consider Case~II, with the boundary condition of a vanishing
lateral flux at $\breve x=\breve l$. Evidently, the above procedure remains
essentially intact, yet with the replacement of the expression for
$\varepsilon$ in Eq.~\eqref{eq:varepsilon-def} by
$\varepsilon=\frac{\beta^2}{2}\breve x(2\breve l-\breve x)$. Thus, the factor $\breve x(\breve l-\breve x)$ is replaced by $\breve x(2\breve l-\breve x)$.  


\section{On the linear stability problem}
\label{app:linearized_stability}

In this appendix, we provide details on the linear stability formulation of
Sec.~\ref{sec:linear_stability}. Our primary hydrodynamic variables are the
rescaled density $\breve \rho$ and vector-valued flux $(\breve J_x, \breve
J_y)=(\breve \rho \breve v_x, \breve \rho \breve v_y)$.

Consider the $y$-independent governing motion laws of
Eq.~\eqref{eq:scaled-model_with_B}. By multiplying the first (second)
equation by $\breve\rho$ and the third equation by $\breve v_x$ ($\breve
v_y$) and then adding the respective two equations together, we obtain the
system
\begin{align}
  \label{eq:app:Jx}
  \frac{\partial \breve{J}_x}{\partial \breve t} + \frac{\partial}{\partial
  \breve x} \biggl(\frac{\breve{J}_x^2}{\breve\rho}\biggr)
  \;&=\;
  (2b)^{-1}\left(-\frac{1}{2}\frac{\partial\breve \rho^2}{\partial \breve
  x} + \beta\,\breve J_y\right),
  \\
  \label{eq:app:Jy}
  \frac{\partial \breve J_y}{\partial \breve t} + \frac{\partial}{\partial
  \breve x}\biggl(\frac{\breve J_x \breve J_y}{\breve \rho}  \biggr)
  \;&=\;
  -\beta\,\breve J_x,
  \\
  \label{eq:app:mass-conserv}
  \frac{\partial \breve \rho}{\partial \breve t}+ \frac{\partial \breve
  J_x}{\partial \breve x}
  \;&=\; 0.
\end{align}
 We also impose one of two sets of boundary conditions.
For Case~I, one such set comes from Eq.~\eqref{eq:scaled-BCs-opposite}, viz.,
\begin{align}
  \label{eq:app:bcs-rescaled-left}
  \breve\rho &= 1\quad \text{at }\breve x=0,
  \\
  \label{eq:app:bcs-rescaled-right}
  \breve J_x&= 1\quad \text{at }\breve x=\breve l,\\
  \label{eq:app:bcs-rescaled-mixed}
  &\breve J_y\big|_{\breve x=0}+\breve J_y\big|_{\breve x=\breve l}=0.
\end{align}
For Case~II, the boundary conditions are described by Eq.~\eqref{eq:scaled-BCs}, which amounts
to Eqs.~\eqref{eq:app:bcs-rescaled-left}
and~\eqref{eq:app:bcs-rescaled-right} with
\begin{align}\label{eq:app:bcs-rescaled-lateralf-zero}
  \breve J_y=0 \quad \text{at }\breve x=\breve l.
\end{align}

Next, we perturb the governing equations around the steady-state solution
$(\breve J_x^s, \breve J_y^s, \breve \rho^s)$, writing
\begin{align*}
  (\breve J_x, \breve J_y)&=(\breve
  J_x^s(\breve x), \breve J_y^s(\breve x))+(\breve\jmath_1(\breve x),
  \breve\jmath_2(\breve x))e^{-\breve \Lambda \breve t},
  \\
  \breve \rho&=\breve \rho^s(\breve x)+\breve \rho_1(\breve x)e^{-\breve
  \Lambda \breve t},\quad \mbox{for}\ 0<\breve x< \breve l.
\end{align*}
Our purpose now is to linearize the governing equations for the
perturbation $(\breve \jmath_1, \breve \jmath_2, \breve\rho_1)$. This step
is illustrated by the following successive approximations:
\begin{align*}
  &(\breve J_x^s+\breve\jmath_1 e^{-\breve \Lambda \breve t})^2
  \;\simeq\;
  (\breve J_x^s)^2+2\breve J_x^s\,\breve \jmath_1 e^{-\breve \Lambda\breve t},
  \\
  &(\breve \rho^s+\breve\rho_1 e^{-\breve \Lambda \breve t})^{-1}
  \;\simeq\;
  (\breve\rho^s)^{-1} \left(1-\frac{\breve\rho_1}
  {\breve\rho^s}e^{-\breve\Lambda \breve t}\right),
  \\
  &(\breve J_x^s+\breve\jmath_1 e^{-\breve \Lambda \breve t})^2
  (\breve \rho^s+\breve\rho_1 e^{-\breve \Lambda \breve t})^{-1}
  \;\simeq\;
  (\breve \rho^s)^{-1}\\
  &\qquad \times \left\{(\breve J_x^s)^2+ \breve
  J_x^s\left(2\breve\jmath_1 -\breve J_x^s\,\frac{\breve\rho_1}
  {\breve\rho^s}\right)e^{-\breve\Lambda\breve t}\right\},
  \\
  &(\breve\rho^s+\breve\rho_1 e^{-\breve\Lambda \breve t})^2
  \;\simeq\;
  (\breve\rho^s)^2+2\breve\rho^s\,\breve\rho_1 \,e^{-\breve\Lambda \breve t}~,
  \\
  &(\breve J_x^s+\breve\jmath_1 e^{-\breve \Lambda\breve t})(\breve
  J_y^s+\breve\jmath_2 e^{-\breve \Lambda\breve t})
  \;\simeq\;
  \breve J_x^s \breve J_y^s \\
  & \qquad +(\breve\jmath_1 \breve J_y^s+\breve \jmath_2
  \breve J_x^s) e^{-\breve \Lambda \breve t},
  \\
  &(\breve \rho^s+\breve\rho_1 e^{-\breve \Lambda \breve t})^{-1}(\breve
  J_x^s+\breve\jmath_1 e^{-\breve \Lambda\breve t})(\breve
  J_y^s+\breve\jmath_2 e^{-\breve \Lambda\breve t})
  \;\simeq\;
  (\breve \rho^s)^{-1}\\
  &\qquad \times \left\{ \breve J_x^s \breve J_y^s + \left(\breve
  \jmath_1 \breve J_y^s+\breve \jmath_2 \breve J_x^s-\breve J_x^s \breve
  J_y^s\frac{\breve\rho_1}{\breve\rho^s}\right)e^{-\breve\Lambda \breve
  t}\right\}.
\end{align*}
The simplification of the nonlinear terms in
Eqs.~\eqref{eq:app:Jx}--\eqref{eq:app:mass-conserv}  by virtue of  these
approximate expressions and further manipulations yield
Eq.~\eqref{eq:eigenvalue_problem} for $\breve \Lambda=\im \breve \omega$.
 The homogeneous boundary conditions of
Eq.~\eqref{eq:eigenvalue_problem_boundary_conditions}
or~\eqref{eq:eigenvalue_problem_boundary_conditions-alt} for
$(\breve\jmath_1, \breve \jmath_2, \breve \rho_1)$ follow from the fact
that the steady-state solution, $(\breve J_x^s, \breve J_y^s, \breve
\rho^s)$, satisfies boundary conditions~\eqref{eq:app:bcs-rescaled-left}
and~\eqref{eq:app:bcs-rescaled-right} with
Eq.~\eqref{eq:app:bcs-rescaled-mixed}
or~\eqref{eq:app:bcs-rescaled-lateralf-zero}.


\section{Numerical approach}
\label{app:numerical_approach}
In this appendix, we outline our numerical approach for solving   the
eigenvalue problems of Cases~I and II.   The algorithms
have been implemented in the Julia programming language
\cite{bezanson2017julia}. 

\subsection{Finite-difference approximation}

We discretize the operator of Eq.~\eqref{eq:eigenvalue_problem} with a
central difference quotient under elliptic regularization. For a given
number $N$ of subintervals of $[0,\breve l]$, for the finite-difference
approximation we set $\breve x_i = \breve h\,i$ for $i=0$, \ldots $N$,
where $\breve h = \breve l / N$. We also introduce finite difference
stencils approximating the first and second derivatives of an arbitrary
discrete function $f(\breve x_i)$ as follows:
\begin{gather*}
  \delta_{\breve h} f_i = \frac{f_{i+1} - f_{i-1}}{2\breve h}, \;
  \delta_{\breve h}^+ f_i = \frac{f_{i+1} - f_{i}}{\breve h}, \;
  \delta_{\breve h}^- f_i = \frac{f_{i} - f_{i-1}}{\breve h},
  \\[0.25em]
  \Delta_{\breve h} f_i = \frac{f_{i+1} - 2f_i + f_{i-1}}{\breve h^2}.
\end{gather*}
Here, we employ the short notation $f_i:=f(\breve x_i)$. Note that the
central-difference stencils $\delta_{\breve h} f_i$ and $\Delta_{\breve h}
f_i$ are second-order approximations of the first and second derivative of
$f$, whereas $\delta^+_{\breve h} f$ and $\delta^-_{\breve h} f$ lead to
first-order approximations.   Let $\mathbb{C}$ denote the set of all complex numbers.   The desired discrete eigenvalue problem
then reads as follows: Find a nontrivial $\overline U^{\breve
h}\in\mathbb{C}^{N+1}$ and $\breve\omega^{\breve h}\in\mathbb{C}$ such that
for $i=1,\ldots,N-1$ we have
\begin{gather*}
  \breve h^2 c \,\Delta_{\breve h} \overline U^{\breve h}_i \;+\;
  \delta_{\breve h} \Big\{\mathbb{A}(\breve x_i) \overline U^{\breve h}_i \Big\}
  \;+\; \beta \mathbb{B}\,\overline U^{\breve h}_i
  \;=\; \im\,\breve\omega^{\breve h}\, \overline U^{\breve h}_i.
\end{gather*}
The factor $c\in\{-1,1\}$ is set to $1$ for approximating the subsonic
spectrum ($\delta<0$) and to $-1$ for the supersonic spectrum ($\delta>0$);
thus, $c=-\mathrm{sgn}(\delta)$ where $\mathrm{sgn}$ is the signum
function. The system is supplemented with three additional difference
equations on the left and right boundaries where we have to take a
one-sided difference operator: We introduce two equations with
forward-difference stencil for the first two components ($k=1,2$) at
$\breve x_0$, the left boundary; and one equation with backward-difference
stencil for the last component ($k=3$) at $\breve x_N$, the right boundary.
These equations read
\begin{align*}
  \delta^+_{\breve h} \Big\{\big(\mathbb{A}(\breve x_0) \overline U^{\breve h}_0\big)_{k} \Big\}
  + \beta \big(\mathbb{B}\,\overline U^{\breve h}_0\big)_{k}
  &= \im\,\breve\omega^{\breve h}\, \big(\overline U^{\breve h}_0\big)_{k},
  \\[0.25em]
  \delta^-_{\breve h} \Big\{\big(\mathbb{A}(\breve x_N) \overline U^{\breve h}_N\big)_{k} \Big\}
  + \beta \big(\mathbb{B}\,\overline U^{\breve h}_N\big)_{3}
  &= \im\,\breve\omega^{\breve h}\, \big(\overline U^{\breve h}_N\big)_{3}.
\end{align*}
Note that for each component we enforce the equation with a one-sided
difference operator on the \emph{opposite side} to the boundary where we
have to enforce a corresponding Dirichlet boundary condition on the
component. 

 Finally, to apply the requisite boundary conditions, we complete the linear system
by imposing three additional equations. Two of these equations are
\begin{align*}
  42^2 \big(\overline U^{\breve h}_N\big)_{1}
  = \im\,\breve\omega^{\breve h} \big(\overline U^{\breve h}_N\big)_{1},
  \qquad
  42^2 \big(\overline U^{\breve h}_0\big)_{3}
  = \im\,\breve\omega^{\breve h} \big(\overline U^{\breve h}_0\big)_{3}.
\end{align*}
Furthermore, depending on whether we enforce
Eq.~\eqref{eq:eigenvalue_problem_boundary_conditions} or
Eq.~\eqref{eq:eigenvalue_problem_boundary_conditions-alt} for the tangential flux, under Case~I or~II, respectively, we impose one of the following equations:
\begin{align*}
  \begin{aligned}
    &\text{Case I:} &\quad
    \frac{42^2}{\breve h^2}\,
    \big(\overline U^{\breve h}_0\,+\,\overline U^{\breve h}_N\big)_{2}
    \;&=\; \im\,\breve\omega^{\breve h}\, \big(\overline U^{\breve h}_N\big)_{2}.
    \\[0.25em]
    &\text{Case II:} &\quad
    42^2\, \big(\overline U^{\breve h}_N\big)_{2}
    \;&=\; \im\,\breve\omega^{\breve h}\, \big(\overline U^{\breve
    h}_N\big)_{2}.
  \end{aligned}
\end{align*}

The numerical value $42^2$ shown above is guaranteed not to coincide with
any of the eigen-frequencies $\breve\omega^{\breve h}$. For
 Case I, this scheme implies that
\begin{align*}
  \overline U^{\breve h}_0+\overline U^{\breve h}_N=0+\mathcal{O}(h^2).
\end{align*}
Correspondingly, for Case II we have $\overline U^{\breve h}_N=0$.

We can now collect all equations into a \emph{stiffness matrix} $\mathbb S$
where $\mathbb S \in \mathbb{R}^{(N+1)\times(N+1)}$. The final system of
linear equations thus reads
\begin{align}
  \label{eq:eigenvalue_problem_implemented}
  \mathbb{S}_{\breve h} \overline U^{\breve h}
  \;=\;
  \im\, \breve\omega^{\breve h}
  \,\overline U^{\breve h}.
\end{align}

\subsection{Implementation and Richardson extrapolation}
The discrete eigenvalue problem described by
Eq.~\eqref{eq:eigenvalue_problem_implemented} is implemented in the
\emph{Julia} programming language.
Due to the fact that Eq.~\eqref{eq:eigenvalue_problem_implemented} expresses a
non-symmetric eigenvalue problem of modest size, we have opted to apply
Julia's \texttt{eigen()} function which internally uses \texttt{LAPACK}
routines for computing the spectrum and eigenvectors. This approach is
reasonably efficient for problem sizes up to $N\leq2500$, which is our
target resolution. A possible alternative suitable for larger systems would
have been a \emph{Krylov subspace method} such as the Krylov-Schur
algorithm.

We have verified numerically that we are indeed in an \emph{asymptotic
regime} for increasing degree of resolution with $N=320$, $640$, $1280$,
and $2560$. This means that the observed numerical error decreases with $N$
with a convergence order of $2$. In order to minimize the error, we have
opted  to perform a single step of a \emph{Richardson extrapolation},
exploiting the fact that we are in an asymptotic regime.  In this vein, we
first compute the spectrum for $N=1280$ and $N=2560$. Then, we filter the
spectrum to relevant eigen-frequencies and identify all corresponding
eigen-frequencies $(\breve\omega_n^{\breve h}, \breve\omega_n^{\breve
h/2})$ between the two sets, and extrapolate as follows:
\begin{align*}
  \breve\omega_n^{\text{extr.}}
  \;=\;
  \frac{\breve\omega_n^{\breve h/2}-2^{-2}\breve\omega_n^{\breve h}} {1-2^{-2}}.
\end{align*}
The resulting numerical value, $\breve\omega_n^{\text{extr.}}$, is
fourth-order convergent. We estimate the final relative discretization
error to be less than $2\%$.

 
\section{On limit of fluid spectrum as $\beta\to 0$}
\label{app:limit-zero-B}

In this appendix, we study the limit of the fluid spectrum as the magnetic
field tends to zero. We examine Cases~I and~II separately.

Consider Eq.~\eqref{eq:eigenvalue_problem}. The  coefficients of this system are well behaved for
sufficiently small $|\bB|$. Hence, we can obtain the limit of the solution
$(\overline U(\breve x), \breve \omega)$ as $\beta\to 0$
by setting $\beta=0$ in
Eq.~\eqref{eq:eigenvalue_problem}. The resulting system reads
\begin{gather}
  \label{eq:app:eigenvalue_problem-zeroB}
  \mathbb{A}^{\!0} \frac{\partial}{\partial \breve x}
  \overline U^{(0)}(\breve x) \;=\; \im\, \breve\omega^0\, \overline U^{(0)}(\breve x),
\end{gather}
where $\overline U^{(0)}=(\breve \jmath_1^{\,0}, \breve \jmath_2^{\,0},
\breve\rho_1^{\,0})^T$ and
\begin{gather*}
  \mathbb{A}^{\!0}=
  \begin{pmatrix}
    2 & 0 & \frac{1-2b}{2b}
    \\[0.5em]
    0 & 1 & 0
    \\[0.5em]
    1 & 0 & 0
  \end{pmatrix}.
\end{gather*}
We seek nontrivial solutions $(\overline U^{(0)}, \breve\omega^0)$ under 
Eq.~\eqref{eq:eigenvalue_problem_boundary_conditions} for Case~I; or, alternatively,
Eq.~\eqref{eq:eigenvalue_problem_boundary_conditions-alt} for Case~II.

The eigenvalues of matrix $\mathbb{A}^{\! 0}$ are $\lambda_0=1$ and
$\lambda_{\pm}=1\pm (2b)^{-1/2}$. The corresponding eigenvectors are
generated from $\overline U_{0}^{(0)}=(0, \lambda_0, 0)^T$ and $\overline
U^{(0)}_{\pm}=(\lambda_\pm, 0, 1)^T$. Hence, the general nontrivial
solution to Eq.~\eqref{eq:app:eigenvalue_problem-zeroB} reads
\begin{equation}
\overline U^{(0)}(\breve x)=\sum_{p=0,\pm}C_p e^{\im (\breve \omega^0/\lambda_p)\breve x}\, \overline U^{(0)}_{p},
\end{equation}
where $C_p$ ($p=0,\,\pm$) is an arbitrary complex number with $(C_0, C_+,
C_-)\neq (0,0,0)$. This formulation yields 
\begin{align}
  \breve \jmath_1^{\,0}(\breve x)&=\sum_{p=\pm}C_p \lambda_p
  e^{\im\tfrac{\breve \omega^0}{\lambda_p}\breve
  x},
  \label{eq:app:j1-zeroB} \\
  \breve \jmath_2^{\,0}(\breve x)&=C_0 e^{i\tfrac{\breve \omega^0}{\lambda_0} \breve x},
  \label{eq:j2-zeroB}\\
  \breve \rho_1^{\,0}(\breve x)&=\sum_{p=\pm}C_p  e^{\im\tfrac{\breve
  \omega^0}{\lambda_p}\breve x},\qquad 0\le \breve x\le \breve l.
  \label{eq:app:rho1-zeroB}
\end{align}
The eigen-frequencies $\breve\omega^0$ must now be determined by imposition
of the boundary conditions. Note the linear independence of the lateral-flux field from
the other two scalar fields.

\medskip 
\subsection{Zero lateral flux at the right boundary (Case~II)}
\label{subsec:app:Zero-tang-flux}
We apply Eq.~\eqref{eq:eigenvalue_problem_boundary_conditions-alt}.
Consequently, by $\breve \jmath_2^{\,0}(\breve l)=0$ we obtain $C_0=0$ for
any $\breve\omega^0$. This in turn implies that $\breve
\jmath_2^{\,0}(\breve x)$ {\em vanishes everywhere}. The constants $C_\pm$
satisfy
\begin{align}
  \label{eq:app:C+-_sys-DS}
  \begin{rcases*}
    \begin{aligned}
      C_+ + C_- &=0,
      \\
      C_+ \lambda_+ e^{\im\tfrac{\breve \omega^0}{\lambda_+}\breve
      l}+C_-\lambda_- e^{\im\tfrac{\breve \omega^0}{\lambda_-}\breve l}&=0.
    \end{aligned}
  \end{rcases*}
\end{align}
By virtue of $(C_+, C_-)\neq (0, 0)$, we derive the whole DS spectrum for
$\breve \omega^0$, Eq.~\eqref{eq:DS-eigenv-instab}. Thus, in this case the
DS predictions are uniquely recovered as $\beta\to 0$ in our model. The normal flux and number density are nontrivial.

\subsection{Opposite lateral fluxes at boundaries (Case~I)}
\label{subsec:app:Opp-tang-flux}
Let us now apply Eq.~\eqref{eq:eigenvalue_problem_boundary_conditions},
which includes the condition $\breve \jmath_2^{\,0}(0)=-\breve
\jmath_2^{\,0}(\breve l)$. Accordingly, we obtain
\begin{equation}\label{eq:app:C0-cond}
  C_0\cos\big(\tfrac{\breve\omega^0\breve l}{2}\big)=0.
\end{equation}
In addition, Eq.~\eqref{eq:app:C+-_sys-DS} must hold. Thus, we need to
distinguish two cases for $C_0$ (and $\breve\omega^0$). First, if $C_0=0$ then we
obtain the DS solution with the familiar DS spectrum and nonzero normal flux and number density. 

On the other hand, if
$C_0\neq 0$, we find 
\begin{equation}\label{eq:app:alt_spec-zeroB}
  \breve\omega_n^0= (2n+1)\frac{\pi}{\breve l}\qquad (n\in \mathbb{Z})
\end{equation}
and
\begin{align}
  \breve \jmath_2^{\,0}(\breve x)=C_0 e^{\im (2n+1) \tfrac{\pi}{\breve
  l}\breve x},\quad 0\le \breve x\le \breve l.
\end{align}
Accordingly, we assert that $C_+=C_-=0$ which yields {\em zero density and normal flux everywhere}. 
In conclusion, in Case~I the total fluid spectrum consists of the DS spectrum and the extraneous set of real-valued eigen-frequencies from
Eq.~\eqref{eq:app:alt_spec-zeroB}.

\end{document}